\documentstyle[12pt,a4wide,epsfig]{article}

\title{
\vspace*{-2.8cm}
\begin{flushright}
{\normalsize DO--TH 96/04 \\ OHSTPY-HEP-T-96-005\\}
\end{flushright}
\vspace{0.8cm}
{\boldmath $B\rightarrow J/\psi\,X$ \unboldmath 
\bf in the Inclusive Parton \\[-1.5mm]and the ACCMM Model}
\thanks{ This work was supported in 
part by the Bundesministerium f\"ur Forschung und Technologie, 056DO93P(5), 
Bonn, FRG, by the CEC Science Project $\mbox{n}^{o}$ SC1-CT91-0729,
and by the US Department of Energy under grant DOE/ER/01545-605} 
}
\author{ 
William F. Palmer\\[-1mm]
Department of Physics, The Ohio State University\\[-1mm]
Columbus, OH 43210, USA\\ \\[-1mm] 
Emmanuel A. Paschos, Peter H. Soldan \\[-1mm] 
Institut f\"ur Physik, Universit\"at Dortmund \\[-1mm]
D-44221 Dortmund, Germany 
}
\date{February 22, 1996}
\begin{document}
\maketitle
\begin{abstract}
The parton model as  
developed for semileptonic $B$ decays is 
applied to the inclusive decay $B\rightarrow J/\psi\,X$.
We calculate the momentum spectrum of the $J/\psi$ using a one-parameter 
distribution function for the heavy quark, taken from production
experiments, and compare our results with recent data from CLEO which fixes 
the distribution parameter $\varepsilon_p$.
An analogous calculation is carried out in the ACCMM model where 
the data determines the Fermi motion parameter $p_f$.

The models give a good description of the data, provided in each of them
the parameter in the distribution function is chosen to correspond to 
a softer $b$ quark momentum distribution than that commonly used
in studies of semileptonic $B$ decays. In particular we arrive at 
$\varepsilon_p={\cal O}(0.008)$ and $p_f={\cal O}(0.55$ GeV), respectively.
The latter value ($p_f=0.5$ GeV) produces in the ACCMM model
$10^2\times |V_{ub}/V_{cb}|^2=1.03$, which removes a former
supposed discrepancy with the ISGW model.
Finally, the strength of the effective color singlet coefficient is found 
to be $|a_2|={\cal O}(0.28)$ in both models.
\end{abstract}
\newpage
\section{Introduction}

The distribution of charmonium states in $B$ meson decays is of special
interest, because it provides a testing ground for color suppression
in $B$ decays already attainable in the present experiments. At the $b$ mass
scale, a leading order calculation generates
a consistent pattern of color suppression for the production of
$S$-wave charmonium states. In the context of the `color singlet mechanism'
the production is related to the decay of a $b$ quark
in the $B$ meson, at short distances, into
a color singlet $c\bar{c}$ pair plus other quarks and gluons. The $c$
and $\bar{c}$ have almost equal momenta and reside in the appropriate
angular-momentum state. This subprocess is dominated by the color suppressed
internal $W$-exchange diagram, where hard gluon exchanges lead to
an effective neutral current \cite{Bodwin}.
Deviations from this leading order `singlet mechanism', which arise from
relativistic corrections and soft gluon induced fragmentation of the
$c\bar{c}$ into charmonium, affect the normalization but not the
structure of the amplitude $b\rightarrow J/\psi\,X$
(in which $X$ sums over hadronic states) \cite{Ko}.

Several authors made predictions for the branching ratio of the direct decay
$b\rightarrow J/\psi\,X$ to leading order in $\alpha_s$ [1--5].
Recently a next-to-leading order QCD calculation was reported \cite{Berg}.
The authors pointed out that the process under consideration cannot 
be explained presently by a standard application of perturbative QCD.
The difficulties entering 
the analysis arise from the strong suppression
of the leading order color singlet Wilson coefficient, causing
considerable cancellations in three different orders of $\alpha_s$.
Alternatively one may use a phenomenological factorization
approach and take into account deviations from this prescription by 
replacing the color singlet renormalization coefficient by an effective 
neutral current coefficient $a_2$, which has to be determined from
experimental data in a model dependent way.

Both ARGUS and CLEO reported on inclusive $B$ decays to $J/\psi$,
where they identified a sizable component of decays with three or more 
particles in the final state \cite{Argus1, Cleo3}. In a recent publication 
\cite{Cleo1} CLEO presented an analysis based on a data sample,
which is one order of magnitude larger than those of previous studies,
corresponding to an error reduction by a factor of $2.4\,$. A comparison 
between theory and experiment
requires the branching ratio resulting from {\it direct} decays. 
Therefore the group corrected the measured $J/\psi$ spectrum for the 
`feed-down' modes $B\rightarrow \psi(2s)\,X$ and $B\rightarrow \chi_{c1}\,X$. 
As a result they found the direct branching ratio $(0.80\pm 0.08)\%$. 
A theoretical analysis 
carried out by the CLEO group following Ref.~\cite{Bodwin} yields $0.75\%$,
in good agreement with the measurement, where the phenomenological 
factorization prescription has been applied using a value $|a_2|=0.23\,$.

The latter value has to be modified when incorporating
bound state effects for the initial $B$ meson, which have to be included
in order to gain a theoretical prediction 
of the $J/\psi$ momentum spectrum within an inclusive approach.
So far there is not available a detailed
fit to experimental data of momentum spectra obtained from theory.
Palmer and Stech have made a preliminary attempt using a simple 
wave function formalism \cite{Pal-St}.
In this paper we will investigate two different approaches in order to 
account for the experimental results.

As stated above a sizable component of the momentum spectrum is due to
nonresonant multi-particle final states. Consequently, for a wide range of
phase-space an inclusive description using quark-hadron duality,
which was extensively applied in the inclusive {\it semileptonic} decays of
$B$ mesons, will be appropriate.

The semileptonic decays involve large energies and
momentum transfers of the weak current over most of the phase-space.
Several groups noticed that these kinematic regions investigate the
light-cone behaviour of currents for which the methods of deep inelastic
scattering are valid. There are various approaches available which
derived useful results describing the lepton spectrum in the decay 
$B\rightarrow X_{u(c)}e\nu$.  
These approaches use different formalisms, but they still follow two basic 
steps.\\[3mm]
(i) The decay of the $b\rightarrow u(c)$ which are
propagating as free particles is considered and the resulting spectrum 
is folded with the momentum distribution of the $b$ quark obtaining moments 
of the quark distribution \cite{Jin1, Jin2}. 
A similar approach is given by the ACCMM model \cite{Alta}, in which one 
accounts for the binding effects by treating the spectator quark as particle 
on the mass shell with nonzero momentum and averages over this momentum.
\\[3mm]   
(ii) After the introduction of the point-like behaviour  
one still has to calculate the expectation value of 
the bilocal transition operator between the initial hadronic state to gain 
the width of the decay. 
In this context the new development is the application of the operator 
product expansion (OPE), which involves an expansion in inverse powers of the 
heavy quark mass $m_b$ incorporating the formalism of the Heavy Quark 
Effective Theory (HQET) \cite{HQET}. 
The calculation of the lepton energy spectra requires a modified
expansion in powers of $1/(1-y)m_b$ with $y$ being the normalized lepton 
energy [18--23].\\[3mm]
The analytic form of the structure function of the heavy quark
within the $B$ meson controls the endpoint behaviour 
of the leptonic spectrum, since in this region the OPE does not converge.
It has been shown that the prediction of HQET far from the endpoint gives
approximately the same shape of the spectrum as the ACCMM model with 
a Fermi motion parameter $p_f\simeq 0.3$ GeV \cite{Bigi4}. 
Yet the model dependence of this region is quite small \cite{Lisa}. Therefore  
it is of great relevance to test the approaches with decay channels of 
the $B$ meson different from the semileptonic one in order to extract the
momentum distribution function of the heavy quark in a direct way and then
compare it with the distribution function obtained in semileptonic decays.

The inclusive decay $B\rightarrow J/\psi\,X$ provides a well suited
testing ground since, as we shall show, in this process 
the dependence on the structure function of the $b$ quark is more direct, 
i.e., the $J/\psi$ momentum spectrum is proportional
to the distribution function.

A QCD based analysis of the decay using HQET is only of limited 
validity, since due to
the smaller energy release the convergence of the OPE is less reliable 
than for semileptonic decays. Therefore one has to resort to a
formalism in which the structure function can be modeled in terms of 
parameters that may be obtained from experiment.
We will apply two different approaches. In Section \ref{IPM-sec} we
present a field-theoretical version of the inclusive parton model, which 
two of us together with Jin 
already applied for semileptonic decays \cite{Jin1, Jin2}. 
In Section \ref{AC-sec} we
calculate the $J/\psi$ momentum spectrum within the framework of the ACCMM
model. In each of the analyses we determine the distribution parameter
of the structure function from a comparison to the recent CLEO data.
The overall normalization of the theoretical spectra further provides
information respecting the value of the effective color singlet coefficient
$|a_2|$. 
Finally we use our results to extract the value of $|V_{ub}/V_{cb}|$ 
in the ACCMM model from the semileptonic decay channel of the $B$. 
This procedure requires some additional remarks which are given    
in Section \ref{remarks}. In Section \ref{polarization} we
compare our results with experimental data for the polarization
of the $J/\psi$. The summary can be found in Section \ref{summary}.

\section{\boldmath $B\rightarrow J/\psi\,X$ \unboldmath
in the Inclusive Parton Model (IPM) \label{IPM-sec}}
\subsection{Calculation of the Differential Branching Ratio}

In order to investigate the interaction which induces 
charmonium production in $B$ decays, we start at the quark level at a 
high-energy scale where the Hamiltonian is known. 
This is renormalized to lower energies to produce
an effective four-fermion interaction
\begin{equation}
{\cal H}_{ef\hspace{-0.5mm}f}=\frac{G_F}{\sqrt{2}}V^{ }_{cb}V_{cs}^*
\left[ \left(c_2+\frac{1}{3}c_1\right) \bar{s}\gamma_\mu^L b\,\bar{c}
\gamma^\mu_L c
+\frac{1}{2}c_1\bar{c}\gamma_\mu^L\lambda^i c\,\bar{s}\gamma^\mu_L \lambda^i 
b\right]+\{s\rightarrow d\} \, ,   \label{Leff}
\end{equation}
$\gamma_\mu^L=\gamma_\mu (1-\gamma_5)$, 
where operators arising from penguin and box diagrams have been neglected. The
first term is a color singlet, the second a color octet operator. The 
renormalization functions (Wilson coefficients) $c_i(\mu)$ have been computed
up to the next-to-leading order corrections in Ref.~\cite{Buras}.

The effective interaction (\ref{Leff}) is evaluated using a factorization
prescription by which the amplitude can be written as a product of matrix
elements of current operators.
Deviations from this prescription are parametrized by substituting the
color suppressed singlet coefficient by a free parameter $a_2$ which is
written in terms of the effective number of colors $(1/\xi)$ and has to be
determined by model-dependent fits to the experimental data,
\begin{equation}
\left(c_2+\frac{1}{N_c}c_1\right)\hspace{5mm} \rightarrow \hspace{5mm}
a_2=c_2+\xi c_1 \, .
\end{equation}
$a_2$ is equivalent to the coefficient introduced for type-II processes 
in the factorization model of BSW \cite{Wirbel} for {\it exclusive} decays.   
Using CLEO data from two-body decay modes with $J/\psi$ mesons 
in the final state, an analysis within the BSW model yields \cite{a2exp}
\begin{equation}
|a_2|=0.26\pm 0.01 \pm 0.01 \pm 0.02 \, , \label{a2}
\end{equation}
where the second systematic error is due to the $B$ meson production 
fractions and lifetimes. Theoretical uncertainties mainly due to hadronic
form factors are not included. A corresponding analysis \cite{Browder}
using the model of
{\it Neubert et al.} \cite{Neubert} provides a central value $|a_2|=0.23$, 
whereas that of {\it Deandrea et al.} \cite{Deandrea} 
yields $|a_2|=0.25$ (for $|V_{cb}|=0.041$, $\tau_B=1.44$ ps and 
$f_{D(D^*)}=220$ MeV).

In this paper we extend the factorization model for exclusive two-body 
decays to an inclusive picture at the quark level. This is reasonable
because, as was stated in Ref.~\cite{Waldi},
many-body final states will most likely start
as two color singlet quark antiquark pairs, including intermediate massive 
resonances, where strong phases from final state interactions disappear in
the sum of all states. The authors of Ref.~\cite{Waldi} made use of this
inclusive picture to determine the lifetime ratio $\tau(B^+)/\tau(B^0)$.
As we intend to reproduce not only the branching ratio but also the
momentum spectrum for the decay $B\rightarrow J/\psi\,X$,
we have to incorporate the momentum distribution of the $b$ quark in the
$B$ meson. In our analysis we will let the effective color singlet 
coefficient $|a_2|$ to float within the range which is covered by the
various models for exclusive modes.

We consider the matrix element of the Hamiltonian (\ref{Leff}) for our
process (in the explicit notation we only refer to the dominant decay
involving an $s$ quark in the final state),  
\begin{equation}
\langle J/\psi\, X_s| {\cal H}_{ef\hspace{-0.5mm}f} |B\rangle=
a_2 \frac{G_f}{\sqrt{2}}V^{ }_{cb}V_{cs}^*\langle J/\psi|\bar{c}
\gamma_\mu^L c|0\rangle \langle X_s| \bar{s}\gamma^\mu_L b|B\rangle\, ,
\label{mel}
\end{equation}
where we make use of the phenomenological factorization prescription. The
second term of Eq.~(\ref{Leff}) does not contribute between color
singlet states.

For the matrix element of the $\bar{c}c(1S)$ at the origin
we use the current-identity for pure vector-like states,
\begin{equation}
\langle 0|\bar{c}(0)\gamma_\mu^L c(0)|J/\psi\rangle =
\varepsilon_{\mu} f_\psi M_{\psi} \, .  \label{ci}
\end{equation}
This identification is valid, because the time-scale of the interaction
to be considered is larger than the scale for the formation of the
singlet state $(t_{int}>1/M_\psi$). $f_\psi$ is the
$J/\psi$ decay constant and can, to leading order in the relative
heavy quark velocity $v$, be related
to the radial part of its nonrelativistic wave function. In principle
one can determine $f_\psi$ from the electromagnetic decay 
$J/\psi\rightarrow e^+ e^-$, but practically one is
confronted with the ignorance of higher order QCD 
corrections to the decay rate which might be of great relevance in view
of existing large leading order corrections. 
Within the studies of {\it exclusive} two-body decays
$B\rightarrow J/\psi\,M$ \cite{Neubert} the central value of the
decay constant was fixed at
$f_\psi=0.384$ GeV which results from the omission of any QCD correction
(as of relativistic corrections to the wave function of the $J/\psi$).
This analysis refers to the 1990 Particle Data Group value for the branching
ratio of $J/\psi\rightarrow e^+ e^-$.
Therefore when extracting $|a_2|$ from {\it inclusive} $B$ decays we
will use the above value in order to compare the result with Eq.~(\ref{a2}).
This procedure requires some additional remarks which concern both the studies
of exclusive and inclusive decays.

Relating the parameter which contains the nonperturbative effects in
the production of $1S$-charmonium
with the $J/\psi$ decay constant implies the neglect
of color octet contributions to the fragmentation process in the explicit
calculation. In Ref.~\cite{Ko} it was argued in the framework of nonrelativistic
QCD (NRQCD), that the fragmentation of a $(\bar{c}c)_8(^3S_1)$ state
into a $J/\psi$ in the long-distance scale may give sizable contributions
to the $B\rightarrow J/\psi\,X$ decay rate. This feature occurs, despite the
fact that the color octet operator is of the order ${\cal O}(v^4)$
respecting the velocity counting rules in the NRQCD, because of the large
suppression of the color singlet Wilson coefficient.

As the structure of the amplitude for the effective point-like decay
$b\rightarrow J/\psi\,X$ remains unchanged, considering nonleading
effects in the fragmentation process only affects the normalization of the
$B\rightarrow J/\psi\,X$ decay spectrum. Using the leading order value
$f_{\psi}=0.384$ GeV therefore means that the effective neutral current
coefficient $|a_2|$, which is determined from a comparison with experimental
data, includes contributions due to the charmonium fragmentation.
Consequently, since these contributions are different for various
charmonium states \cite{Ko}, the universality of $|a_2|$ is limited to $B$
decays involving fixed $S$-wave resonances.\\ \\
Reducing the matrix element (\ref{mel}) of the mesonic decay $B\rightarrow
J/\psi\,X_s$ to that of the point-like effective neutral current subprocess
$b\rightarrow J/\psi\,s$ (see Fig.~1c) in the valence quark approximation
one obtains
\begin{equation}
\langle J/\psi\, X_s|{\cal H}_{ef\hspace{-0.5mm}f}|B\rangle=a_2 V^{ }_{cb}
V_{cs}^*M_{\psi}f_{\psi}\frac{G_F}{\sqrt{2}}\varepsilon^*_{\mu}
\bar{s}\gamma^{\mu}(1-\gamma_5)b \, .
\end{equation}
This expression for a free quark decay has to be modified by incorporating
bound state corrections due to the strong interaction between heavy and
light quark in the initial meson. 
In the {\it intuitive} parton model one has to take the 
incoherent sum over all subprocesses in which a distribution function $f(x)$ 
is introduced as a weight,
\begin{equation}
\Gamma_B = \int_0^1 dx\, f(x) \Gamma_b(x) \, , \hspace{5mm} 
\mbox{where} \hspace{5mm}\int_0^1 f(x)\,dx=1 
\, . \label{ipm}
\end{equation}  
In Eq.~(\ref{ipm}) transverse momenta of the heavy quark relative to
the B meson are neglected which leads to the Lorentz invariant prescription
$p_b^\mu = x P_B^\mu$. This approach corresponds to an equal
velocity approximation, i.e., valence quarks and $B$ meson have the
same velocity, and is valid at the infinite momentum frame. We therefore
proceed as usual computing a Lorentz invariant quantity
and then use it in any other frame.

The basis for our approach is given by a {\it field-theoretical} version
of the parton model. If we square the second matrix element on the 
right-hand side of Eq.~(\ref{mel}) and sum over all final states, which
guarantees incoherence, we produce the hadronic tensor corresponding to the 
effective transition 
$B\rightarrow X_f$,
\begin{equation}
W_{\mu\nu}=-\frac{1}{2\pi}\int d^4 y \,e^{iqy}\langle B| \left[
j_\mu(y),\, j_\nu^\dagger (0)\right]|B\rangle \, ,
\end{equation}   
where $j_\mu(x)=\,:\hspace{-1mm}\bar{q}_f\gamma_\mu^L b(x)\hspace{-1mm}:$ is
the left-handed effective neutral current.
The same tensor structure we encountered in the semileptonic
$B$ meson decays [11--14].
Substituting for the leptonic tensor the corresponding
expression of the $J/\psi$ current,
\begin{equation}
L_{\mu\nu}(J/\psi)=2\pi^3 |V_{cf}|^2 |a_2|^2 f_\psi^2 M_{\psi}^2
\left(-g_{\mu\nu}+\frac{(k_\psi)_\mu (k_\psi)_\nu}{M_\psi^2}\right)\, ,
\label{psiten}
\end{equation}
we can write the differential rate for the decay $B\rightarrow X_f\,J/\psi$
in the restframe of the $B$ in exact analogy to the semileptonic decay,
\begin{equation}
d\Gamma_{(B\rightarrow X_f\,J/\psi)}=\frac{G_F^2 |V_{cb}|^2}{(2\pi)^5M_B}
L_{\mu\nu}W^{\mu\nu}\frac{d^3k_\psi}{2E_\psi}\,.
\end{equation} 
The general structure of the hadronic tensor reads
\begin{eqnarray}
W^{\mu\nu}&=&-g^{\mu\nu}W_1+\frac{1}{M_B^2}P_B^\mu P_B^\nu W_2-i\varepsilon^{
\mu\nu\alpha\beta}\frac{1}{M_B^2}P_{B\alpha}q_\beta W_3
\label{tensor1} \\
&& +\frac{1}{M_B^2}q^\mu q^\nu W_4
+\frac{1}{M_B^2}(P_B^\mu q^\nu+P_B^\nu q^\mu)W_5+\frac{1}{M_B^2}
i(P_B^\mu q^\nu-P_B^\nu q^\mu)W_6 \,, \nonumber
\end{eqnarray}
where in our case $q=k_\psi$.
Introducing the light-cone dominance as in Ref.~[11--15] allows to refer
the tensor corresponding to the transition $B\rightarrow X_f$ in the
decay $B\rightarrow X_f\,J/\psi$
to the distribution function $f(x)$ of the heavy quark momentum,
\begin{equation}
W_{\mu\nu ,f}=4(S_{\mu\rho\nu\lambda}-i\varepsilon_{\mu\rho\nu\lambda})
\int dx\,f(x)P_B^\lambda(xP_B-k_\psi)^\rho \varepsilon[(xP_B-k_\psi)_0]
\delta[(xP_B-k_\psi)^2-m_f^2]\, ,
\label{tensor2}
\end{equation} 
with 
\begin{equation}
S_{\mu\rho\nu\lambda}=g_{\mu\rho}g_{\nu\lambda}-g_{\mu\nu}g_{\rho\lambda}
+g_{\mu\lambda}g_{\nu\rho} 
\hspace{5mm} \mbox{and} \hspace{5mm}
\varepsilon (x)=\left\{ \begin{array}{ll}
   +1\, , \hspace{4mm} & x > 0 \\ -1 \, , & x < 0 \, . \end{array}  \right.  
\end{equation} 
{}From Eqs.~(\ref{tensor1}) and (\ref{tensor2})  one obtains the structure
functions $W_i$ of the hadronic tensor in the restframe of the $B$ meson,
which two of us together with Jin derived in Ref.~\cite{Jin1, Jin2}
in exact analogy for the semileptonic decay rate,
\begin{eqnarray}
\begin{array}{lcl}
W_1=2[f(x_+)+f(x_-)]\, ,& \hspace{4mm} & \displaystyle
W_2=\frac{8}{x_+-x_-}\left[x_+ f(x_+)-x_- f(x_-)\right]\, ,\\[4mm]
\displaystyle W_3=W_5=\frac{-4}{x_+-x_-}\left[f(x_+)-f(x_-)\right]\, ,& 
\hspace{3mm} & W_4=W_6=0\, , 
\label{wi}
\end{array}
\end{eqnarray}
where we have defined
\begin{equation}
\hspace*{-2mm}
M_B x_\pm=\frac{1}{M_B}\left(P_B k_{\psi}\pm
\sqrt{(P_Bk_{\psi})^2+(m_f^2-M_{\psi}^2)M_B^2}\,\right)
\stackrel{(|\vec{p}_B|=0)}{=}\left(E_{\psi}\pm\sqrt{
|\vec{k}_{\psi}|^2+m_f^2}\,\right) \, .
\end{equation}
For the case of a massless final state quark $x_\pm$ are identical to
the usual light-cone variables. 
The dependence of the distribution function on the single
scaling variable $x$ is a consequence of the light-cone dominance, since 
in this framework the
structure function $f(x)$ is obtained as the Fourier transform of the
reduced bilocal matrix element, which contains the long-distance 
contributions to the hadronic tensor \cite{Jin3},
\begin{equation}
f(x)=\int d(yP_B)e^{ix(yP_B)}\frac{1}{4\pi M_B^2}
\langle B|\bar{b}(0)P_B^\mu \gamma_\mu^L b(y)|B\rangle|_{y^2=0} \, .
\label{disfunc}
\end{equation} 
The terms proportional to $f(x_-)$ in Eq.~(\ref{wi}) are a result of the
field-theoretical approach. The kinematical range for $x_-$ belongs to
a final state quark with negative energy. Therefore the corresponding terms
can be associated, formally, with
quark pair-creation in the $B$ meson (see Fig.~1b), whereas the dominant
terms proportional to $f(x_+)$ reflect the direct decay of Fig.~1a.

Including the small $f(x_-)$ term as well as the CKM-suppressed 
transition $c\rightarrow d$ the Lorentz invariant width 
of the decay $B\rightarrow J/\psi\,X$ reads
\begin{eqnarray}
E_B\cdot d\Gamma_{(B\rightarrow J/\psi\,X)}
&=&\sum_{f=s,\,d}\frac{|C_f|^2}{2\pi^2}\int dx\, f(x)
\left[ P_B(xP_B-k_\psi)+\frac{2}{M_\psi^2}(P_Bk_\psi)(xP_B k_\psi-M_\psi^2)
\right] \nonumber \\
&& \times \varepsilon[(xP_B-k_\psi)_0]\delta^{(1)}\left[(xP_B-k_\psi)^2-m_f^2
\right]\frac{d^3k_\psi}{2 E_\psi} \,  
\label{diffbr}
\end{eqnarray}
with
\begin{equation}
|C_f|=\frac{G_F}{\sqrt{2}}|V_{cb}||V_{cf}|M_{\psi}f_{\psi}|a_2|\,.
\label{cf}
\end{equation} 
Here the modification from the field-theoretical approach simply enters in 
form of the step-function $\varepsilon(x)$ which exactly provides the 
additional terms $f(x_-)$ in the tensor coefficients (\ref{wi}). 
Evaluating Eq.~(\ref{diffbr}) in the restframe of the $B$ meson we arrive at
the formula for the $J/\psi$ momentum spectrum,
\begin{equation}
\frac{d\Gamma_{(B\rightarrow J/\psi\,X)}}{d|\vec{k}_{\psi}|}= 
\sum_{f=s,\,d}\frac{|C_f|^2}{4\pi M_B}\left[ 3W_1+W_2\frac{|\vec{k}_\psi|^2}
{M_\psi^2}\right]\frac{|\vec{k}_\psi|^2}{E_\psi}, 
\end{equation}
within the framework of the inclusive parton model.
Using Eq.~(\ref{wi}) the corresponding differential branching ratio can be
written as 
\begin{eqnarray}
\lefteqn{\frac{1}{\Gamma_B}
\frac{d\Gamma_{(B\rightarrow J/\psi\,X)}}{d|\vec{k}_{\psi}|}
\left(|\vec{k}_{\psi}|,\,|\vec{p}_B|=0 \right)= } 
\label{br}\\
&&\tau_B\sum_{f=s,\,d}\frac{|C_f|^2}
{2\pi M_B}\frac{|\vec{k}_{\psi}|^2}{E_{\psi}}\left[f(x_+)
\left(1+\frac{2E_{\psi}M_B}{M_{\psi}^2}\left(x_+-\frac{2 m_f^2}{M_B^2
(x_+-x_-)}\right)\right)+(x_+\leftrightarrow x_-)\right] \, ,
\nonumber
\end{eqnarray}
within the kinematical range  
\begin{equation}
M_{\psi}\le E_{\psi}\le \frac{M_B^2+M_{\psi}^2-S_{min}^{(f)}}{2M_B} \, ,
\hspace{1cm} S_{min}^{(f)}=m_f^2 \, .
\label{kin}
\end{equation}
To compare our result (\ref{br}) with data from CLEO a Lorentz boost has  
in addition to be performed from the restframe of the $B$ meson
to a $B$ produced at the $\Upsilon (4S)$ resonance ($|\vec{p}_B|=0.34$ GeV).
The explicit form of the boost integral we give in Section \ref{ac-cal}
in the context of the ACCMM model (see Eq.~(\ref{brac})).\\ \\
All terms are known in this decay except for the product $(|a_2|f_\psi)$
and the structure function which occurs with two arguments. We note that
when neglecting the small $f(x_-)$ contribution the decay spectrum is directly
proportional to $f(x_+)$. Thus measuring the momentum spectrum of the $J/\psi$
we can read from the data the distribution function. The extraction of 
$f(x_+)$ is direct, in contrast to the extraction from the electron energy
spectrum in the semileptonic decay of the $B$ meson which involves an integral
over the structure function.

In the following Section we compare the predictions of Eq.~(\ref{br})
with the existing experimental data.
\subsection{Analysis and Numerical Evaluation \label{subnum1} }

The probability distribution function (\ref{disfunc}) is not Lorentz invariant
and is defined in the infinite momentum frame. 
It cannot be transformed to the restframe of the $B$
meson (or a frame which corresponds to a $B$ meson produced at the 
$\Upsilon (4s)$ resonance), because it involves an infinite sum of 
quark-antiquark pairs whose calculation requires a complete solution of the 
field-theory. In the absence of direct measurements of the distribution function
we use a one-parameter Ansatz and fix the distribution parameter
by comparing our results with data. Referring to theoretical studies
which pointed out that the distribution and fragmentation function of heavy
quarks peak at large values of $x$ \cite{Bjorken, Brodsky},
as a working hypothesis we assume
that the functional form of both is similar. The latter is known from
experiment and we shall use the Peterson functional form [35--37]
\begin{equation}
f_\varepsilon(x)=N_\varepsilon
\frac{x(1-x)^2}{[(1-x)^2+\varepsilon_p x]^2}\, ,
\label{peterson}
\end{equation}
with $\varepsilon_p$ being the free parameter and  
$N_\varepsilon$ the corresponding normalization constant.
This form we already applied in the semileptonic decays of the $B$ meson 
\cite{Jin1, Jin2}.

In Fig.~2a we show the distribution function $f_\varepsilon(x)$
for various values of $\varepsilon_p$. 
The kinematical range for the two arguments $x_\pm(m_f)$ of the distribution
function according to Eq.~(\ref{kin}) reads
\begin{equation}
\frac{M_{\psi}+m_f}{M_B}\le x_+\le 1\, , \hspace{1cm} 
\frac{M_{\psi}^2-m_f^2}{M_B^2}\le x_-\le \frac{M_{\psi}-m_f}{M_B} \, .
\end{equation}
As pointed out in Fig.~3 the variable $x_-$ only occurs with values at which
$f_\varepsilon (x_-)$ is small. Therefore the corresponding contribution
to the differential branching ratio (\ref{br}) is small.

We use the distribution (\ref{peterson}) to fit the measured momentum spectrum 
of the $J/\psi$ which was given by the CLEO group \cite{Cleo1}. 
As argued above we fix $f_\psi=0.384$ GeV and vary $|a_2|$ according to
the range covered by the various models for exclusive decays.

The shape of the theoretical spectrum is determined by the distribution
parameter $\varepsilon_p$, which is illustrated in Fig.~4 where
we show the spectrum for $|a_2|=0.275$ and several values of $\varepsilon_p$.  
A general feature of the analysis is our 
difficulty in reproducing the data over the whole range of phase-space.  
Confronted with this problem, we lay greater emphasis on the 
appropriate description of the {\it low} momentum range 
($|\vec{k}_\psi| \le 1.4$ GeV) in our simultaneous fits of $|a_2|$, which in
this decay has the meaning of a pure normalization constant, and the 
distribution parameter $\varepsilon_p$. 
 Within this region the $J/\psi$ spectrum 
obtains a sizable contribution from decay channels containing 
three or more particles in the final state (where higher $K^*$ resonances
are assumed to be unimportant), whereas the high momentum range  
is purely determined by the exclusive two-body decays 
$B\rightarrow J/\psi\,K^{(*)}$. Therefore incoherence as a necessary 
ingredient of the IPM is limited to the former region.

Furthermore we demand the reproduction
of the measured branching ratio for the decay under consideration.  
Thus if we apply small values for the distribution parameter 
$\varepsilon_p\le 0.006$ we cannot account for the low momentum region,
which is underestimated significantly within the theoretical spectrum
(see Fig.~5).
On the other hand, demanding the reproduction of the branching ratio, a 
distribution corresponding to $\varepsilon_p \ge 0.010$
would imply a value $|a_2|\ge 0.30$, which is beyond the range
given from the study of exclusive two-body decays.
Therefore, in spite of the difficulty to reproduce the data
accurately over 
the whole range of phase-space, the investigation of the decay 
$B\rightarrow J/\psi\,X$ is more restrictive with regard to the distribution 
parameter than the semileptonic decay of the $B$ meson which involves an 
integral over the structure function.

Considering the fact that the IPM does not include hadronization effects 
and consequently yields an averaged spectrum,
a satisfactory description of the dataset is achieved for  
$\varepsilon_p={\cal O}(0.008)$, 
$|a_2|={\cal O}(0.285)$ (compare Fig.~5) when using current
masses for the final state quarks. Applying constituent masses would involve
an incorrect position of the maximum of the momentum spectrum and, in 
addition, a sizable underestimate of the high momentum range (see
Fig.~6).
Yet also for $\varepsilon_p=0.008$ there is apparent a moderate systematic 
underestimate of the latter region. Since the model presumes the existence 
of multiple final states it 
may not hold for the high momentum range ($|\vec{k}_\psi| \ge 1.4$ GeV)
where two-body decays determine the spectrum.
Following the CLEO analysis, as a trial
we subtract the expected contribution of the exclusive decays
$B\rightarrow K^{(*)}J/\psi$ from the data for the semi-inclusive momentum
spectrum. Carrying out the fit for this modified momentum distribution 
we observe that the agreement between theory and experiment is 
improved for large values of the distribution parameter $\varepsilon_p\ge
0.012$ (however only when applying constituent masses for the final state
quarks), corresponding to a soft momentum distribution of the heavy quark. 
In Fig.~7 we show the modified spectrum for $|a_2|=0.275$ and
$\varepsilon_p$ as stated there. This additional fit may serve as an 
indication that a value $\varepsilon_p={\cal O}(0.014)$ has to be 
suggested within the model when restricting the analysis of decay 
spectra to the `incoherence region' where multi-particle final states
exist. Then additional information from a model describing the dominant
two-body decay channels has to be implemented to reproduce the data
over the whole range of phase-space.

Finally, the errors in the data, although substantially improved, are still
significant and a crucial test will be possible, when the error bars are further
reduced. It is of special interest to establish whether the deviations at
$|\vec{k}_\psi|\simeq 0.5$ GeV from the smooth shape of the spectrum, which
is predicted in our inclusive approach, survive and to extend the analysis of
the composition, in terms of resonances and continuum, for $|\vec{k}_\psi|
\ge 1.4$ GeV.
%
%

\section{\boldmath $B\rightarrow J/\psi\,X$ \unboldmath in the ACCMM Model
\label{AC-sec}}

\subsection{Calculation of the Differential Branching Ratio \label{ac-cal}}

A second approach which allows us to determine the momentum distribution
of the $J/\psi$ in the semi-inclusive decay of the $B$ meson is given through
the ACCMM model \cite{Alta}. 
In this model the bound state corrections to the
simple quark picture are incorporated by attributing to the spectator
quark a Fermi motion within the meson. The momentum spectrum of the 
$J/\psi$ is then obtained by folding the Fermi motion with the spectrum from
the $b$ quark decay. In Ref.~\cite{Barger}
the shape of the $J/\psi$ momentum distribution resulting from Fermi
momentum smearing has been given in the restframe of the $B$ meson,
but without a comparison with data, and
without consideration of the light spectator quark mass. 
Moreover the authors
applied the constituent mass for the strange quark in the final state,
whereas we 
will show that a satisfactory reproduction of the data can only be 
obtained applying the current mass.

According to Eq.~(\ref{ci}) the $J/\psi$ is treated as a color singlet
current using the factorization assumption of Eq.~(\ref{mel}).
The spectator quark is handled as on-shell particle with definite mass 
$m_{sp}$ and momentum $|\vec{p\hspace{0.2mm}}|=p$.
Consequently, the $b$ quark is considered to be off-shell with a
virtual mass $W$ given in the restframe of the $B$ meson
by energy-momentum conservation as
\begin{equation}
W^2(p)=M_B^2+m_{sp}^2-2M_B\sqrt{m_{sp}^2+p^2}\,. \label{vmass}
\end{equation}
{\it Altarelli et al.} introduced in the model a gaussian probability
distribution $\phi(p)$ for the spectator (and thus for the heavy 
quark) momentum, 
\begin{equation}
\phi(p)=\frac{4}{\sqrt{\pi}p_f^3}\exp\left(-p^2/p_f^2\right) \, , 
\label{ac-dist}     
\end{equation}
normalized according to 
\begin{equation}
\int_0^\infty dp\, p^2 \phi(p) = 1\, .
\end{equation}
Here a free parameter $p_f$ is adopted for the gaussian width which has to
be determined by experiment.

The main difference between the inclusive parton model and ACCMM is that 
the latter one must consider a $b$ quark in flight.   
We therefore start from the momentum spectrum of the $J/\psi$ 
resulting from the decay 
$b\rightarrow q_f\,J/\psi$ $(f=s,\,d)$ 
of a $b$ quark of mass $W$ and momentum $p$ which is given by     
\begin{equation}
\frac{d\Gamma_b^{(f)}}{d|\vec{k}_{\psi}|}\left(|\vec{k}_{\psi}|,\,p\right)=   
\gamma_b^{-1}
\frac{\Gamma_0^{(f)}}{k_+^{(b,\,f)}(p)-|k_-^{(b,\,f)}(p)|}\left[\theta 
\left(|\vec{k}_{\psi}|-|k_-^{(b,\,f)}(p)|\right)-\theta 
\left(|\vec{k}_{\psi}|-k_+^{(b,\,f)}(p)\right)\right] \, .
\label{gbp}
\end{equation}
Here we have defined 
\begin{equation}
\theta (x)=\left\{ \begin{array}{ll}
   1, \hspace{4mm} & x > 0 \\ 0, & x < 0 \, . \end{array}  \right. 
\end{equation}
$\Gamma_0^{(f)}$ is the width of the analogous decay in the restframe of the
heavy quark where we have confined ourselves to the leading order result,
\begin{equation}
\Gamma_0^{(f)} =
\frac{|C_f|^2}{2\pi}\frac{k_0^{(f)}}{W^2}\left[m_f^2
+\frac{1}{2}\left(W^2-m_f^2
-M_{\psi}^2\right)\left(2+\frac{W^2-m_f^2}{M_{\psi}^2}\right)\right] \, ,
\label{g0}
\end{equation}
with $|C_f|$ according to Eq.~(\ref{cf}) and with the momentum $k_0^{(f)}$ of
the $J/\psi$,
\begin{equation}
k_0^{(f)}=\frac{1}{2W}\left[\left(W^2-m_f^2+M_{\psi}^2\right)^2-4W^2 
M_{\psi}^2\right]^{\frac{1}{2}} \;, \hspace{5mm} E_0^{(f)} = 
\sqrt{k_0^{(f)\,2}+M_{\psi}^2} \, .
\end{equation}
For a vanishing mass $m_f$ of the final quark Eq.~(\ref{g0}) is
equivalent to Eq.~(5) in Ref.~\cite{Wise}.

In Eq.~(\ref{gbp}) $k^{(b,\,f)}_\pm$ give the limits of the momentum range
which results from the Lorentz boost from the restframe of the $b$ 
quark to a frame where the $b$ has a nonvanishing momentum $p$,
\begin{equation}
k^{(b,\,f)}_{\pm}(p)=\frac{1}{W}\left(E_b k_0^{(f)}\pm p E_0^{(f)} \right)\, ,
\end{equation}
and $\gamma_b^{-1}$ is the corresponding Lorentz factor, 
\begin{equation}
\gamma_b^{-1}=\frac{W}{E_b}\;, \hspace{1cm} E_b=\sqrt{W^2+p^2}\,.
\end{equation}
To calculate the momentum spectrum of the $J/\psi$ from the 
semi-inclusive decay of the $B$ meson one has to fold the heavy quark momentum 
probability distribution with the spectrum (\ref{gbp}) resulting from 
the quark subprocess. Performing this
we finally arrive at the expression for the differential branching ratio
for a $B$ meson in flight,  
\begin{eqnarray}
\lefteqn{\hspace*{-1cm}\frac{1}{\Gamma_B}
\frac{d\Gamma_{(B\rightarrow J/\psi\,X)}}{d|\vec{k}_{\psi}|}
\left(|\vec{k}_{\psi}|, \, |\vec{p}_B|\right)= }\label{brac}\\
&&
\tau_B\sum_{f=s,\,d}\hspace{2mm} 
\int\limits_{|k_-(|\vec{k}_{\psi}|)|}^{k^{(f)}_+(|\vec{k}_{\psi}|)}\;
\frac{d|\vec{k}_{\psi}'|}{k_+(|\vec{k}_{\psi}'|)-|k_-(|\vec{k}_{\psi}'|)|}
\int\limits_0^{p_{max}^{(f)}}dp\;p^2\phi(p)
\frac{d\Gamma_b^{(f)}}{d|\vec{k}_{\psi}|}\left(|\vec{k}_{\psi}'|,\,p\right) \, .
\nonumber
\end{eqnarray}
Here $p_{max}^{(f)}$ is the maximum kinematically allowed value of the quark
momentum $p$, i.e., that which makes $W$ in Eq.~(\ref{vmass}) equal to
$W=m_f+M_{\psi}$,
\begin{equation}
p_{max}^{(f)}=\frac{1}{2M_B}\left[(M_B^2+m_{sp}^2-(m_f+M_{\psi})^2)^2
-4m_{sp}^2M_B^2 \right]^\frac{1}{2} \, .
\end{equation}
The first integration in Eq.~(\ref{brac}) results from the transformation from
the spectrum for a $B$ meson at rest to the spectrum for a $B$ meson in flight,
where 
\begin{equation}
k_{\pm}(|\vec{k}_{\psi}|)=\frac{1}{M_B}\left(E_B |\vec{k}_{\psi}|\pm 
|\vec{p}_B| E_{\psi}\right) \, , \hspace{5mm}
k_+^{(f)}(|\vec{k}_{\psi}|)=\mbox{min}\{k_+(|\vec{k}_\psi|),\, 
k_{max}^{(f)}\}\, , 
\end{equation}
$k_{max}^{(f)}$ being the maximum value of the $J/\psi$ momentum from the 
decay $B\rightarrow J/\psi\,X_f$ in the restframe of the $B$,
\begin{equation}
k_{max}^{(f)}=\frac{1}{2 M_B}\left[\left(M_B^2+M_\psi^2
-S_{min}^{(f)}\right)^2-4 M_B^2 M_\psi^2\right]^{\frac{1}{2}}\, ,\hspace{5mm} 
S_{min}^{(f)}=(m_f+m_{sp})^2\, .  \label{lim}
\end{equation}

In the following Section we make use of Eq.~(\ref{brac}) to compare the
model predictions with experimental data.
\subsection{Analysis and Numerical Evaluation \label{ac-an}}

Both models, the inclusive parton model as well as the ACCMM model
incorporate the bound state structure of the $B$ meson by postulating a
momentum distribution for the heavy quark. 
Introducing in the latter model another $x$-variable as the ratio $x=W/M_B$,
the appropriate distribution function for the relative mass fraction $w(x)$
of the $b$ quark in the restframe of the $B$ meson is given as
\begin{eqnarray}
w(x) & = & \frac{2M_B^2}{\sqrt{\pi}p_f^3}\,p(x)\,x\left(1-x^2+\frac{m_{sp}^2}
{M_B^2}\right)\exp\left[-p(x)^2/p_f^2\right] \, , \nonumber \\[1mm]
p(x) & = & \frac{M_B}{2}\left[(1-x^2)^2-2\frac{m_{sp}^2}{M_B^2}(1+x^2)+
\frac{m_{sp}^4}{M_B^4}\right]^{\frac{1}{2}} \, ,  \label{wx}
\end{eqnarray}
where the corresponding normalization reads
\begin{equation}
\int_0^{1-m_{sp}/M_B}w(x)\,dx=1 \, .
\end{equation}
In Fig.~2b we show the distribution function $w(x)$ for $m_{sp}=0.2$ GeV
and various values of $p_f$. From Eqs.~(\ref{wx}) one realizes
that for given on-shell masses the
Fermi parameter $p_f$ determines the average value 
$\langle W \rangle=M_B \langle x \rangle$
of the virtual $b$ quark mass and hence also the total decay width.

Both models have the advantage of avoiding the mass of the 
heavy quark as an independent parameter. As a consequence, the phase-space is
treated correctly by means of using mesonic degrees of freedom. 
Nevertheless the endpoint of the phase-space in the ACCMM model is slightly
different from that in the IPM because of the on-shell mass $m_{sp}$ of the
spectator quark from which the minimal invariant mass square of the
hadronic system  $S_{min}^{(f)}$ results as given in Eq.~(\ref{lim}).

The shape of the momentum spectrum in the decay $B\rightarrow J/\psi X$ 
is mainly determined by the value of $p_f$. The extraction of this value from
a comparison with experimental data is important not only for
explaining the decay itself and thus testing the factorization 
assumption of Eq.~(\ref{mel}), but also for the determination of
$|V_{ub}/V_{cb}|$ from the endpoint region of the inclusive 
{\it semileptonic} $B\rightarrow X_{c(u)}l\nu$ decay spectrum. As recently
stressed in Ref.~\cite{Kim} the experimental extraction of $p_f$ from
semileptonic decays has been ambiguous, because various parameters of
the model were fitted simultaneously to the
lepton energy spectrum where in addition the perturbative QCD corrections 
are important in the endpoint region
(especially for $b\rightarrow u$).  
Furthermore, as stated in Ref.~\cite{Lisa}, for a large range of
the phase-space
the bound state corrections are of minor importance in this decay channel.

Whereas usually $p_f=0.3$ GeV is used for the data analysis the authors of 
Ref.~\cite{Kim} calculate the Fermi parameter theoretically in the
relativistic quark model and obtain $p_f=0.54$ GeV.
A CLEO analysis \cite{Cleo4} of the endpoint lepton spectrum in the semileptonic
decay channel employing the common value $p_f=0.3$ GeV (and $m_{sp}=0.15$ GeV)
yields a discrepancy between
the ACCMM and the ISGW model of {\it Isgur et al.} \cite{ISGW} in the 
determination of $|V_{ub}/V_{cb}|$, 
\begin{eqnarray}
10^2\times |V_{ub}/V_{cb}|^2 & = & 0.57\pm 0.11 
\hspace{1cm}(\mbox{ACCMM}) \nonumber \\
& = & 1.02\pm 0.20 \hspace{1cm} (\mbox{\it Isgur et al.}) \, .
\end{eqnarray}
Using however $p_f=0.5$ GeV the authors of Ref.~\cite{Kim} arrive at the value
$10^2\times |V_{ub}/V_{cb}|^2=1.03$ within the ACCMM model, 
obtaining a good agreement with the ISGW model.

The decay $B\rightarrow J/\psi\,X$ provides an independent way of
extracting the Fermi parameter $p_f$. In contrast to the lepton 
spectrum in the semileptonic decay the shape of the $J/\psi$
momentum spectrum is highly sensitive to this parameter over the whole
phase-space. This is analogous to the determination of the 
parameter $\varepsilon_p$ in Section \ref{subnum1}$\,$. \\ \\
Employing the spectator distribution function (\ref{ac-dist}) we calculated
the momentum spectrum for the decay of $B$ mesons produced at the 
$\Upsilon(4s)$ resonance. Fig.~8 shows
the comparison with the CLEO data for $f_\psi=0.384$ GeV, $m_{sp}=0.2$ GeV, 
$m_s=0.125$ GeV, $m_d=0$ and various sets of values for $p_f$ and $|a_2|$ 
as stated there. 
Here we want to emphasize that analogous to the IPM the best fit is achieved 
when employing current masses for the final state quarks. This is a consequence
of the fact that within the theory these are treated as free particles
without consideration of fragmentation. The application of constituent masses
would yield a sizable underestimate of the high momentum region (see
Fig.~9).

In addition to the final state quark masses $m_f$ the spectator mass
also determines the range of phase-space and the position of the maximum
(compare Fig.~10). The latter is in accordance with the data for
$m_{sp}=0.2$ GeV.

In Fig.~8 one can see that the agreement of the theoretical
spectrum with the data is very good provided we choose the value 
$p_f={\cal O}(0.55$ GeV).
To demonstrate the high sensitivity of the spectrum on this 
parameter over the whole range of phase-space we present
in Fig.~11 spectra for different values of $p_f$ with $f_\psi$
and $|a_2|$ fixed. It is obvious that the shape of the experimental 
$J/\psi$ momentum distribution  
(compare Fig.~8) cannot be reproduced in the model
when using values for the Fermi motion parameter significantly smaller than  
$p_f=0.5$ GeV. Especially the sizable contribution in the low momentum 
range requires a soft probability distribution of the heavy quark momentum.

As a result, we conclude that the value $p_f={\cal O}(0.55$ GeV) for the
Fermi motion parameter of the ACCMM model is highly favoured when
applying this model to the semi-inclusive decay $B\rightarrow J/\psi\,X$.
This value is in exact accordance with the one calculated recently in 
Ref.~\cite{Kim} from the relativistic quark model. As mentioned above
the discrepancy between
the ACCMM model and the ISGW model concerning the determination of 
$|V_{ub}/V_{cb}|$ disappears when using our favoured value for $p_f$ in the
study of the electron spectrum from inclusive semileptonic $B$ decays.

A final feature of our investigations is related to the determination 
of the effective color singlet coefficient $|a_2|$. 
Using $f_\psi=0.384$ GeV, $\tau_B=1.54$ ps, and $|V_{cb}|=0.043$,
we obtain $|a_2|={\cal O}(0.28)$ for the best fit of the ACCMM model 
to the CLEO data.
This is in agreement to the value of $|a_2|$ which we found in the IPM. 
The study of an inclusive decay compared to exclusive decays
in this context provides the advantage of solely involving 
an analysis that is independent of mesonic form-factors.     
\section{Remarks on the Extraction of
\boldmath $|V_{ub}/V_{cb}|$ \unboldmath  \label{remarks} }

The application of the value for
$p_f$, which we obtain from our investigation of the decay
$B\rightarrow J/\psi\,X$, to the semileptonic decay channel
requires some additional remarks. 
As pointed out in Ref.~\cite{Bigi5} the universality of the momentum
distribution function describing the motion of the $b$ quark in the $B$
meson only holds for different decay processes when referring to a
fixed final state quark mass.

The common procedure, followed in the analysis of the semileptonic decay
$B\rightarrow Xl\nu$, is to determine the distribution parameter from a
fit of the lepton spectrum away from the endpoint where the
spectrum is dominated by $b\rightarrow c$ transitions.
This result is then used to model the endpoint region, which 
purely originates from $b\rightarrow u$ transitions, in order to
extract the value of $|V_{ub}/V_{cb}|$ from the data. Considering however
the fact that the decay $b\rightarrow c$ involves a quark
mass $m_c$, which cannot be neglected, the corresponding 
distribution parameter might be unsuitable to describe 
simultaneously the transition $b\rightarrow u$.

On the other hand, the independent determination of the heavy quark momentum 
distribution from the entire $J/\psi$ spectrum 
of the $B\rightarrow J/\psi\,X$ decay 
involves the effective transition $b\rightarrow s$
(and also the Cabibbo suppressed decay $b\rightarrow d$), 
i.e., a small (current) mass for the final state.
Therefore the corresponding distribution parameter 
is still appropriate for the extraction of $|V_{ub}/V_{cb}|$, 
which requires the distribution function associated with 
the transition to a massless final state.

The CLEO analysis \cite{Cleo4} was performed within the ACCMM model using
the value $p_f=0.3$ GeV for both, $b\rightarrow c$ and $b\rightarrow u$
decays. The modification which arises 
when keeping this value for the transition 
$b\rightarrow c$ but applying $p_f=0.5$ GeV for $b \rightarrow u$ reads
\begin{equation}
 \left|\frac{V_{ub}}{V_{cb}}\right|^2_{p_f=0.5}
=\left|\frac{V_{ub}}{V_{cb}}\right|^2_{p_f=0.3}
 \times \frac{\tilde{\Gamma}(p_f=0.3)}{\tilde{\Gamma}(p_f=0.5)}\, ,
\label{Vub}
\end{equation}
where $\tilde{\Gamma}(p_f)\equiv \int_{2.3}^{2.6} dE_l\frac{d\Gamma}
{dE_l}(p_f)$
denotes the integration over the endpoint domain of the leptonic spectrum
supposing $|V_{ub}|=1$. 
This is exactly the same relation which has been
used in Ref.~\cite{Kim} and we quote the corresponding value for
$|V_{ub}/V_{cb}|^2$ in Section \ref{ac-an}$\,$.
Nevertheless the physical interpretation is somewhat different. 
The authors of Ref.~\cite{Kim} took the value of $|V_{cb}|$ as
determined independently from other analyses. 
This appeared to be necessary, because they did not consider the
limitations to the universality of the distribution function
arising from different masses of the final state quark.
In contrast to this we argue that Eq.~(\ref{Vub}) can be
applied directly in the investigation of the semileptonic spectrum,
since only the distribution parameter governing the endpoint region has to 
be changed relative to the existing CLEO analysis \cite{Cleo4}. 
Within this analysis the application of the ACCMM model to $b\rightarrow c$ 
transitions has to be regarded as a phenomenological fit,
because from a theoretical point of view
the neglect of Fermi motion in the final state is only
appropriate, when small final state quark masses are involved
\cite{Bigi5}.

The latter remark illustrates the importance of a study which is 
independent of the semileptonic data. 
Our investigation of the decay $B\rightarrow J/\psi\,X$
allows a suitable determination of the parameters of the ACCMM model 
which can be used for the analysis of the endpoint 
spectrum in the decay $B\rightarrow Xl\nu$. 
The same statement holds true for our
analysis within the parton model. 

\section{Polarization of the \boldmath $J/\Psi$ \unboldmath
\label{polarization}}

Applying the same method as for the calculation of the unpolarized $J/\psi$
spectrum, we proceed to determine the momentum spectrum of a longitudinally
polarized state.

In references \cite{Kuehn2} and \cite{Pal-St} the polarization of the $J/\psi$
was obtained from a study of the free quark decay $b\rightarrow J/\psi\, s$,
i.e., for fixed momenta in the final state. The result, $\Gamma_L/\Gamma
\simeq 0.54$, was identified with the average polarization in the corresponding
decay of the $B$ meson. Considering the bound state structure of the $B$ within
our inclusive approach yields the momentum distribution of the $J/\psi$.
Consequently, it allows us to determine the polarization in various kinematic
regions, where the results can be compared with measurements from CLEO
\cite{CLpol} and ARGUS \cite{ARpol}.

Since our model is based on local quark-hadron duality, we do not expect to
reproduce the polarization within the region $|\vec{k}_\psi|\ge 1.4$ GeV,
because it is governed by the two-body decay modes
$B\rightarrow J/\psi K^{(*)}$.
As can be concluded from the ARGUS data (see Tab.~1), after subtraction of the
exclusive mode $B\rightarrow J/\psi K$ the high momentum range of the decay
spectrum is dominated by a single orbital angular momentum and consequently
by a single $CP$ eigenstate of the mode $B\rightarrow J/\psi K^*$ \cite{Hon}.
This feature is not accounted for within our inclusive approach since final
state interactions are not part of our consideration.

Nevertheless the study of the low momentum range, which involves nonresonant
multi-particle states, provides an additional test of our approach, especially
in view of future measurements with reduced experimental errors. Furthermore
we investigate the modification of the average polarization
relative to the free quark decay due to the bound state structure of the
$B$, in which the hadronic vertex acts on the polarization of the $J/\psi$
through the momentum distribution of the underlying $b$ quark.\\ \\
Replacing the polarization sum in Eq.~(\ref{psiten}) by
$\varepsilon_\mu (${\small $\lambda =0$})$\varepsilon_\nu
(${\small $\lambda =0$}),
a calculation analogous to the determination of the unpolarized
$B\rightarrow J/\psi\,X$ decay spectrum within the parton model, introduced in
Section \ref{IPM-sec}, yields the momentum spectrum for the longitudinally
polarized $J/\psi$. In the restframe of the $B$,
\begin{eqnarray}
\lefteqn{\frac{1}{\Gamma_B}
\frac{d\Gamma_{(B\rightarrow (J/\psi)_{L}\,X)}}{d|\vec{k}_{\psi}|}
\left(|\vec{k}_{\psi}|,\,|\vec{p}_B|=0 \right)= } \label{polsp}\\
&&\tau_B\sum_{f=s,\,d}\frac{|C_f|^2}
{2\pi M_B}\frac{|\vec{k}_{\psi}|^2}{E_{\psi}}\left[f(x_+)
\left(-1+\frac{2E_{\psi}M_B}{M_{\psi}^2}\left(x_+-\frac{2 m_f^2}{M_B^2
(x_+-x_-)}\right)\right)+(x_+\leftrightarrow x_-)\right] \, ,
\nonumber
\end{eqnarray}
with $|C_f|$ defined according to Eq.~(\ref{cf}).
Note that the difference to the unpolarized spectrum occurs in the first term
in the parenthesis of Eq.~(\ref{polsp}), whose sign changes compared to
Eq.~(\ref{br}). The partial branching ratio in the CLEO frame is obtained by
performing the integration due to the transformation of the spectrum to
that for a $B$ meson in flight (see Eq.~(\ref{brac})), as well as the
integration over the momentum range which is considered.

The calculation in the ACCMM model follows the lines of Section \ref{AC-sec},
where for the polarized case Eq.~(\ref{g0}) is replaced by the corresponding
expression for the decay width  of $b\rightarrow (J/\psi)_L\,q_f$. In the
restframe of the heavy quark one obtains
\begin{equation}
\Gamma_{0,\,L}^{(f)}\, =\,
\frac{|C_f|^2}{2\pi}\frac{k_0^{(f)}}{W^2}\left[-m_f^2
+\frac{1}{2}\left(W^2-m_f^2
-M_{\psi}^2\right)\left(\frac{W^2-m_f^2}{M_{\psi}^2}\right)\right] \, .
\label{g0pol}
\end{equation}
The partial branching ratio for the momentum range
$k_1\le |\vec{k}_\psi|\le k_2$ then reads
\begin{eqnarray}
\lefteqn{\hspace*{-5mm}\Delta B_{[L]}(k_1,\,k_2)=}\label{parpol}\\
& &
\tau_B\sum_{f=s,\,d}\hspace*{2mm}
\int\limits_{k_1}^{k_2} d|\vec{k}_\psi|
\int\limits_0^{p_{max}^{(f)}}dp \; \frac{p^2\phi(p)\gamma_b^{-1}
\Gamma_{0\,[L]}^{(f)}}{k_+^{(b,\,f)}(p)-|k_-^{(b,\,f)}(p)|}
\int\limits_{g_1(|\vec{k}_\psi|,\,p)}^{g_2(|\vec{k}_\psi|,\,p)}\;
\frac{d|\vec{k}_{\psi}'|}{k_+(|\vec{k}_{\psi}'|)-|k_-(|\vec{k}_{\psi}'|)|}
\, ,
\nonumber
\end{eqnarray}
where
\begin{eqnarray}
g_1(|\vec{k}_\psi|,\,p)&=&\mbox{max}\left\{|k_-(|\vec{k}_\psi|)|,\,
|k_-^{(b,\,f)}(p)|\right\} h(|\vec{k}_\psi|,\,p) \, , \\
g_2(|\vec{k}_\psi|,\,p)&=&\mbox{min\hspace{1mm}}
\left\{\hspace*{1.1mm}k_+(|\vec{k}_\psi|)\hspace{1.1mm},\,
\hspace{1.1mm}k_+^{(b,\,f)}(p)\hspace{1.1mm}\right\} h(|\vec{k}_\psi|,\,p) \, ,
\end{eqnarray}
and
\begin{equation}
\label{hlim}
h(|\vec{k}_\psi|,\,p)=
\theta\left[k_+^{(f)}(|\vec{k}_\psi|)-|k_-^{(b,\,f)}(p)|\right]
\theta\left[k_+^{(b,\,f)}(p)-|k_-(|\vec{k}_\psi|)|\right]\, .
\end{equation}
In Eqs.~(\ref{parpol}-\ref{hlim}) the limits of integration are written in a
way as to take into account simultaneously the Lorentz boost from the $b$ to
the $B$ restframe and the subsequent boost to the CLEO frame.

Our results for the average longitudinal polarization of the $J/\psi$,
$\Gamma_L/\Gamma =\Delta B_L/\Delta B$, are presented in Tab.~1.
There we give the polarization obtained in the parton and the
AC\nolinebreak CMM model for different values of the quark momentum distribution
parameters $\varepsilon_p$ and $p_f$, respectively. The $J/\psi$ momentum ranges
that we considered are chosen according to the experimental data from ARGUS
and CLEO.
\begin{table}[thb]
\begin{center}
\begin{tabular}{|ccccc|}
\hline \hline
$J/\psi$ momentum            
& \multicolumn{2}{c}{\hspace*{2.3cm}CLEO II \cite{CLpol}}                           
& \multicolumn{2}{c|}{ARGUS  \cite{ARpol}}    \\ \hline
$k_{\psi}< 0.8$ GeV          
& \multicolumn{2}{c}{\hspace*{2.3cm}$0.55 \pm 0.35$}                          
& \multicolumn{2}{c|}{}               \\
0.8 GeV $<k_{\psi}< 1.4$ GeV 
& \multicolumn{2}{c}{\hspace*{2.3cm}$0.49 \pm 0.32$}                             
& \multicolumn{2}{c|}{}               \\
1.4 GeV $<k_{\psi}< 2.0$ GeV 
& \multicolumn{2}{c}{\hspace*{2.3cm}$0.78 \pm 0.17$}                              
& \multicolumn{2}{c|}{$1.17 \pm 0.17$}\\
all $k_{\psi}< 2.0$ GeV      
& \multicolumn{2}{c}{\hspace*{2.3cm}$0.59 \pm 0.15$}                             
& \multicolumn{2}{c|}{}               \\
\hline 
\multicolumn{5}{c}{}\\[-4mm]
\hline 
\multicolumn{5}{|c|}{\large parton model \rule[-3mm]{0mm}{9mm}}\\ \hline
$J/\psi$ momentum & $\varepsilon_p=0.004$ & $\varepsilon_p=0.006$ & 
$\varepsilon_p=0.008$ & $\varepsilon_p=0.010$ \\ \hline
$k_{\psi}< 0.8$ GeV          & 0.416  & 0.415  & 0.414  & 0.413  \\
0.8 GeV$<k_{\psi}< 1.4$ GeV  & 0.520  & 0.515  & 0.512  & 0.508  \\
1.4 GeV$<k_{\psi}< 2.0$ GeV  & 0.557  & 0.552  & 0.547  & 0.543  \\
all $k_{\psi}< 2.0$ GeV      & 0.537  & 0.529  & 0.522  & 0.516  \\
\hline 
\multicolumn{5}{c}{}\\[-4mm]
\hline
\multicolumn{5}{|c|}{\large ACCMM model \rule[-3mm]{0mm}{9mm}}\\ \hline
$J/\psi$ momentum & $p_f=0.3$ \hspace*{-2mm}& 
$p_f=0.4$  \hspace*{-2mm}& $p_f=0.5$ \hspace*{-2mm}& 
$p_f=0.55$ \hspace*{-2mm}\\ \hline
$k_{\psi}< 0.8$ GeV         &0.515 &0.504 &0.495 &0.491\\
0.8 GeV$<k_{\psi}< 1.4$ GeV &0.548 &0.538 &0.530 &0.527\\
1.4 GeV$<k_{\psi}< 2.0$ GeV &0.556 &0.549 &0.542 &0.539\\
all $k_{\psi}< 2.0$ GeV     &0.553 &0.543 &0.534 &0.529\\
\hline \hline
\end{tabular}
\caption{$J/\psi$ polarization $\Gamma_L/\Gamma $
in the decay$B\rightarrow J/\psi X$ in the parton and the ACCMM model
compared to data. The parameters are $m_{sp}=0.2$ GeV, $m_s=0.125$ GeV, 
$m_d=0$, and $\varepsilon_p$, $p_f[\mbox{GeV}]$ respectively, as given
above.}
\end{center}
\end{table}
{}From the table one can read that in both models the data is reproduced
correctly for the range $|\vec{k}_\psi| \le 1.4$ GeV and also for the
average over the complete momentum range. The agreement is good, but the
experimental errors are still large.
On the other hand, the polarization predicted for the high
momentum region underestimates the data. This is to be expected because, as we
mentioned, the region $|\vec{k}_\psi|>1.4$ GeV is dominated by two-body modes.

Significant differences between the parton and the ACCMM model show up only
for the low momentum range ($|\vec{k}_\psi| \le 0.8$ GeV), which may provide
an additional feature for distinguishing between the models.
The bound state corrections to the average polarization of the $J/\psi$ (for
all $|\vec{k}_\psi| \le 2.0$ GeV) are marginal; a property which is reflected
in the weak dependence of the polarization on the distribution parameters.
As a consequence of the cancellation of binding effects in the ratio
$\Delta B_L/\Delta B$, the determination of $\varepsilon_p$ and $p_f$,
respectively, from the polarization turns out to be impossible.

We conclude that the inclusive approach to $B\rightarrow J/\psi\,X$ decays
in the parton and the ACCMM model yields the polarization of the $J/\psi$ in
various kinematic regions. Our analysis shows that for the high momentum range,
in which final state interactions are important, the approach is less reliable
in the polarized than in the unpolarized case. On the other hand, the inclusive
description is well suited in the low momentum range governed by nonresonant
multi-particle states. A study of the polarization in this range may help
distinguishing between the models considered, as soon as the experimental
error is reduced.
\section{Summary and Conclusion} \label{summary}

We incorporate in this article the bound state effects of the $B$ meson 
in the analysis of the direct decay $B\rightarrow J/\psi\,X$. 
An inclusive approach 
has been worked out in detail within the framework of the parton and 
the ACCMM model where in each case  
a one-parameter momentum distribution function for the heavy quark is introduced.  We fixed the distribution parameter
by comparing the predicted $J/\psi$ momentum spectrum with the recent CLEO 
data, putting emphasis on the adequate reproduction of the sizable low momentum
spectrum, which contains nonresonant multi-particle final states.
Both models yield a $b$ quark momentum distribution softer
than the one commonly used in studies of semileptonic $B$ decays.
Within the parton model we obtain $\varepsilon_p={\cal O}(0.008)$ where
moderate deviations from the Peterson {\it et al.} distribution, taken from
production experiments, are apparent in the data.
In further studies of inclusive $B$ decays within the parton model it will be
of interest to consider a modified Ansatz for the $b$ quark momentum
distribution.

The ACCMM model can account for the data over the whole
range of phase-space when we use a Fermi motion parameter 
$p_f={\cal O}(0.5$ GeV). Applying this result
to the endpoint electron energy spectrum of semileptonic $B$ meson decays
yields a value $10^2\times |V_{ub}/V_{cb}|^2=1.03$, which is
in accordance to the result obtained within the exclusive ISGW model.

The successful reproduction of the experimental data confirms the
validity of the factorization assumption for inclusive $B$ decays where
the bound state effects in both models 
imply a large value for the effective color singlet coefficient, 
$|a_2|={\cal O}(0.28)$ (when choosing $\tau_B=1.54$ ps, $|V_{cb}|=0.043$).
The inclusive approach can also account for the measured average polarization
of the $J/\psi$ which is independent of the normalization constants
$(|a_2|f_\psi)$ and $\tau_B$.

In view of the fact that semileptonic decay spectra
away from the endpoint of the phase-space involve 
$b\rightarrow c$ transitions a  study of 
the decay $B\rightarrow J/\psi\,X$  presently 
provides the only possibility to extract the momentum distribution 
of the $b$ quark corresponding to a decay in which the mass of the quark
in the final state is negligible. This extraction is direct
as no integration over the distribution function has to be performed.
Furthermore the momentum spectrum of the $J/\psi$ is more sensitive 
to the bound state structure of the $B$ than the electron energy spectrum from
semileptonic decays.
The difficulties which arise in the study of a momentum 
spectrum containing resonance structures can be avoided as soon as 
precise measurements are available which allow a determination of the
distribution function from
the endpoint domain of the electron energy spectrum in semileptonic
decays or the photon spectrum in $b\rightarrow s\,\gamma$ decays. 
\vspace*{3cm}\\
\centerline{ \large  Acknowledgements }
\\[0.5cm]
PHS wants to thank the {\it Deutsche Forschungsgemeinschaft}
for financial support (in connection with the Graduate College for
Elementary Particle Physics in Dortmund).\\
\newpage
\vspace*{5cm}
{\large \bf Figure Captions }\\
\begin{itemize}
\item[Fig.~1a]Dominant contribution to the decay
$B\rightarrow J/\psi\,X$ with the light final state quark
$q_f$ ($f=s$, $d$) in the subprocess having positive energy. 
\item[Fig.~1b]Contribution to the decay 
$B\rightarrow J/\psi\,X$ with the light final state quark
$q_f$ ($f=s$, $d$) in the subprocess having negative energy. 
\item[Fig.~1c]$b\rightarrow J/\psi+s(d)$ matrix element. The square 
represents a four-fermion vertex.
\item[Fig.~2a]Momentum (respectively mass) distribution function 
$f_\varepsilon (x)$ in the IPM for various values of $\varepsilon_p$.
\item[Fig.~2b]Mass distribution function $w(x)$ in the ACCMM model for
various values of $p_f$.
\item[Fig.~3 ]$x_\pm (f=s)$ as functions of the
$J/\psi$ momentum for $m_s=0.125$ GeV (current mass)
and for $m_s=0.55$ GeV (constituent mass). 
\item[Fig.~4 ]Theoretical momentum spectrum in the IPM for direct inclusive 
$J/\psi$ production from $B$ decays at the $\Upsilon(4s)$ resonance.
The values for the parameters are $m_s=0.125$ GeV, $m_d=0$
(current masses), $M_B=5.279$ GeV, $f_\psi=0.384$ GeV, $|a_2|=0.275$, 
$\tau_B=1.54$ ps, $|V_{cb}|=0.043$, $|V_{cs}|=0.97$, 
$|V_{cd}|=0.22$ and various values of $\varepsilon_p$ as shown.
\item[Fig.~5 ]Same as in Fig.~4, now theoretical momentum spectrum
for various values of $\varepsilon_p$ and $|a_2|$ as shown, compared with
the CLEO data. 
\item[Fig.~6 ]Same as in Fig.~5, now for $\varepsilon_p=0.008$, $|a_2|=0.285$
and current, respectively constituent masses for the final state quarks. 
\item[Fig.~7 ]Theoretical momentum spectrum in the IPM for direct inclusive 
$J/\psi$ production compared with the corrected CLEO data where the 
expected contribution of
the exclusive two-body final states ($J/\psi \, K^{(*)})$ has been subtracted.
The values for the parameters are $m_s=0.55$ GeV, $m_d=0.33$ GeV,
$|a_2|=0.275$ and $\varepsilon_p$ as shown.
\item[Fig.~8 ]Theoretical momentum spectrum in the ACCMM model for direct 
inclusive $J/\psi$ production from $B$ decays at the $\Upsilon(4s)$ resonance
compared with the CLEO data.
The values for the parameters are $m_s=0.125$ GeV, $m_d=0$
(current masses), $m_{sp}=0.2$ GeV, $M_B=5.279$ GeV,  
$\tau_B=1.54$ ps, $|V_{cb}|=0.043$,  
$|V_{cs}|=0.97$, $|V_{cd}|=0.22$ and various sets of values for $p_f$ 
and $|a_2|$ as shown.   
\item[Fig.~9 ]Same as in Fig.~8 , now for $p_f=0.55$ GeV, $|a_2|=0.28$,
$m_{sp}=0.2$ GeV and current, respectively constituent masses for the 
final state quarks. 
\item[Fig.~10]Theoretical momentum spectrum for direct inclusive $J/\psi$
production in the ACCMM model for $p_f=0.55$ GeV, $|a_2|=0.28$ and various
values of $m_{sp}$ as shown compared with the CLEO data. 
\item[Fig.~11]
Theoretical momentum spectrum 
as in Fig.~10, now for $m_{sp}=0.2$ GeV, $|a_2|=0.28$ and various
values of $p_f$ as shown.
\end{itemize}
%
\newpage
\bf
\large
\noindent
\vspace*{0.4cm}\\
\centerline{fig. 1a}\\[1.4cm]
\noindent
\vspace*{0.4cm}\\
\centerline{fig. 1b}\\[1.4cm]
\noindent
\vspace*{0.4cm}\\
\centerline{fig. 1c}
\newpage
\noindent
\hspace*{1.18cm}\epsfig{file=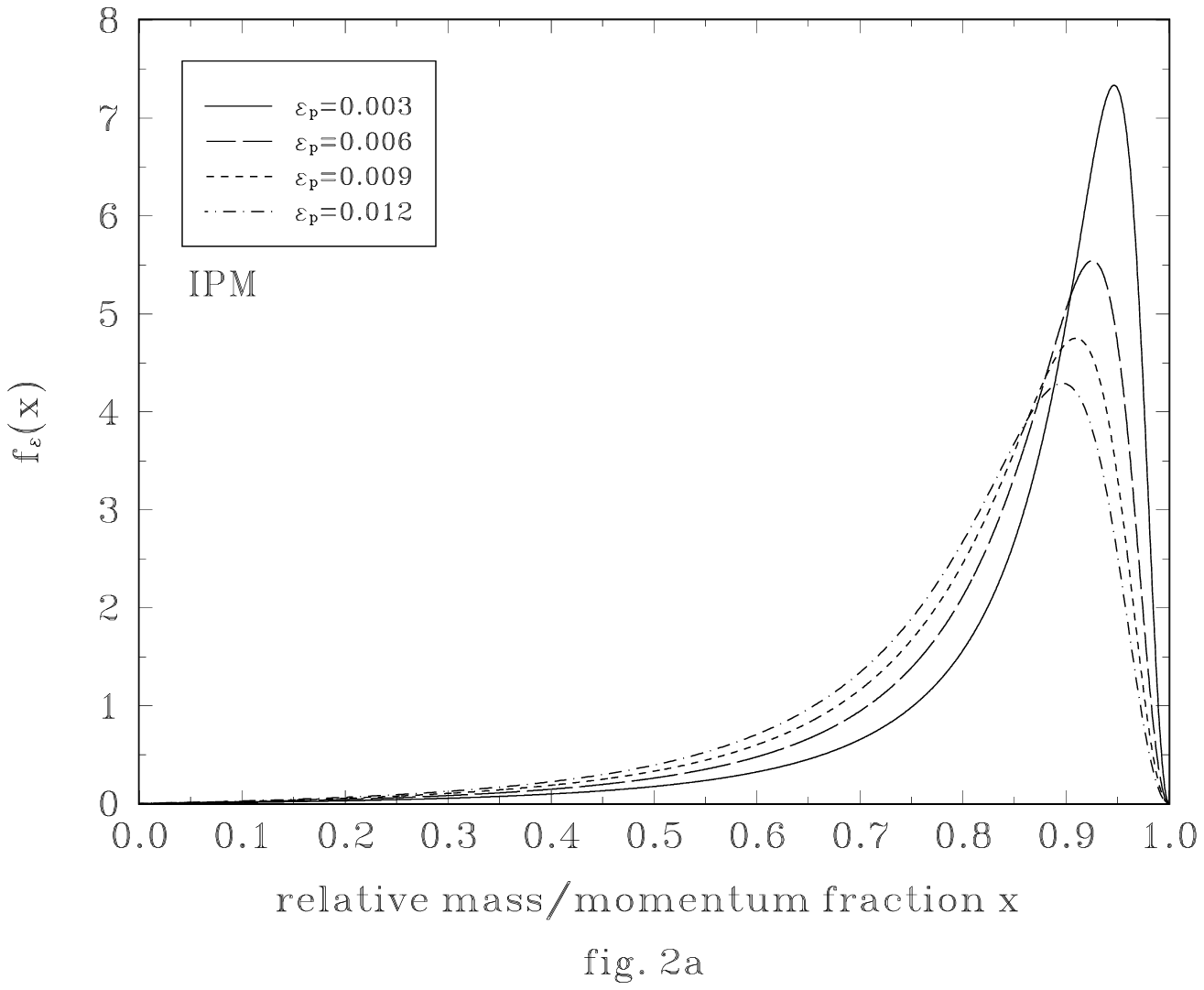,height=11cm,angle=360}\\
\vspace*{-1cm}
\noindent
\hspace*{1cm}\epsfig{file=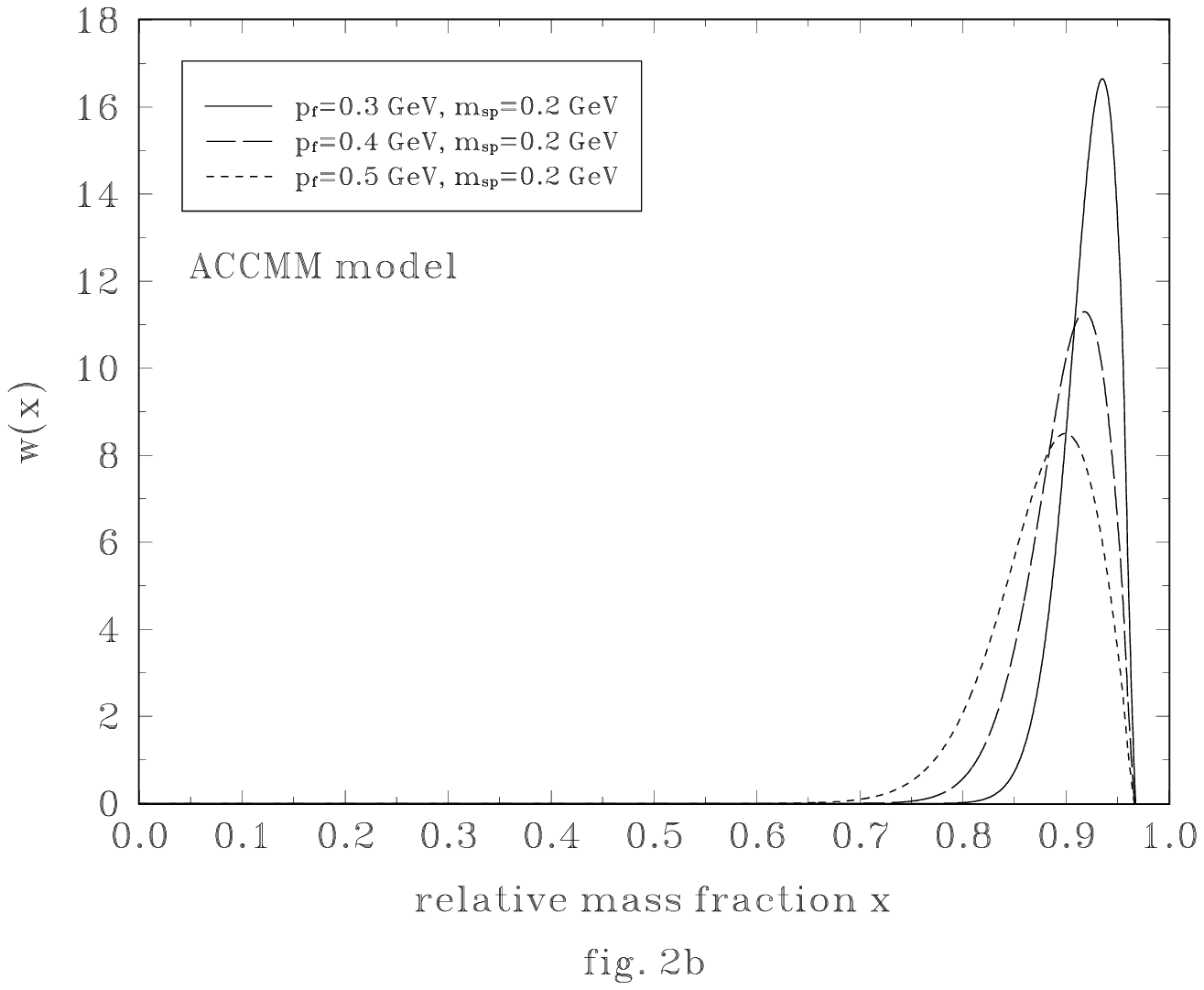,height=11cm,angle=360}
\newpage
\noindent
\hspace*{1.18cm}\epsfig{file=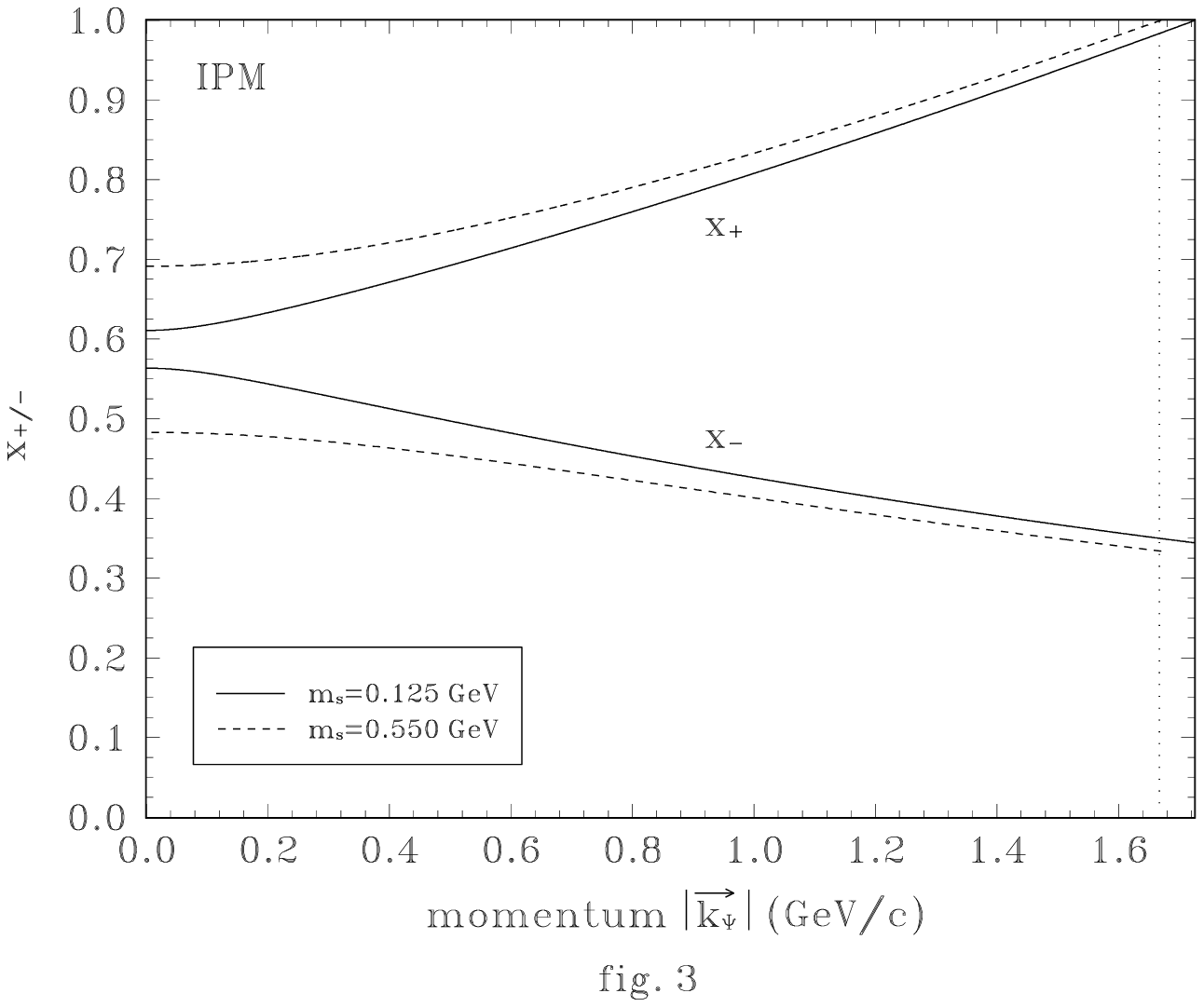,height=11cm,angle=360}\\
\vspace*{-1cm}
\noindent
\hspace*{1cm}\epsfig{file=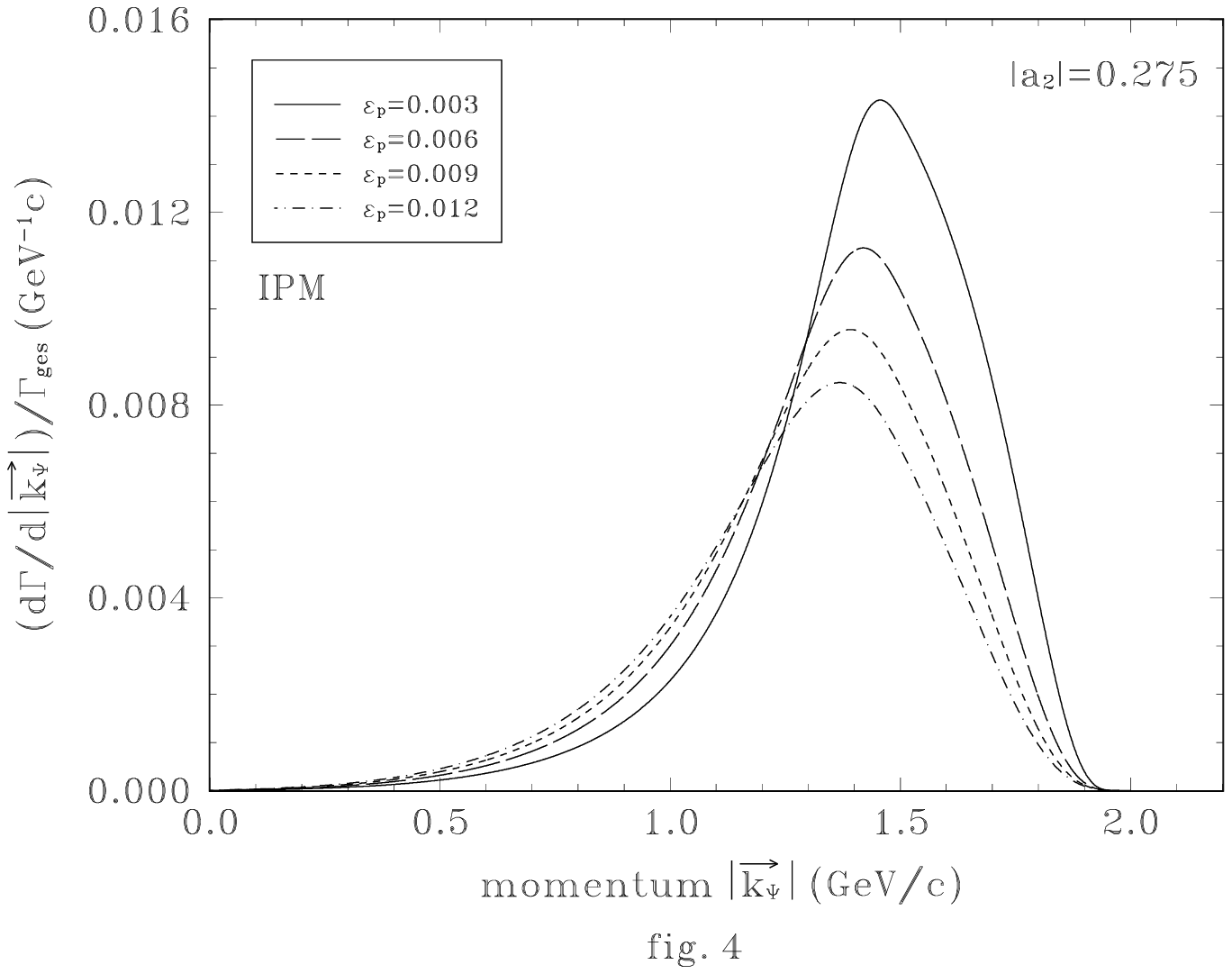,height=11cm,angle=360}
\newpage
\noindent
\hspace*{1.18cm}\epsfig{file=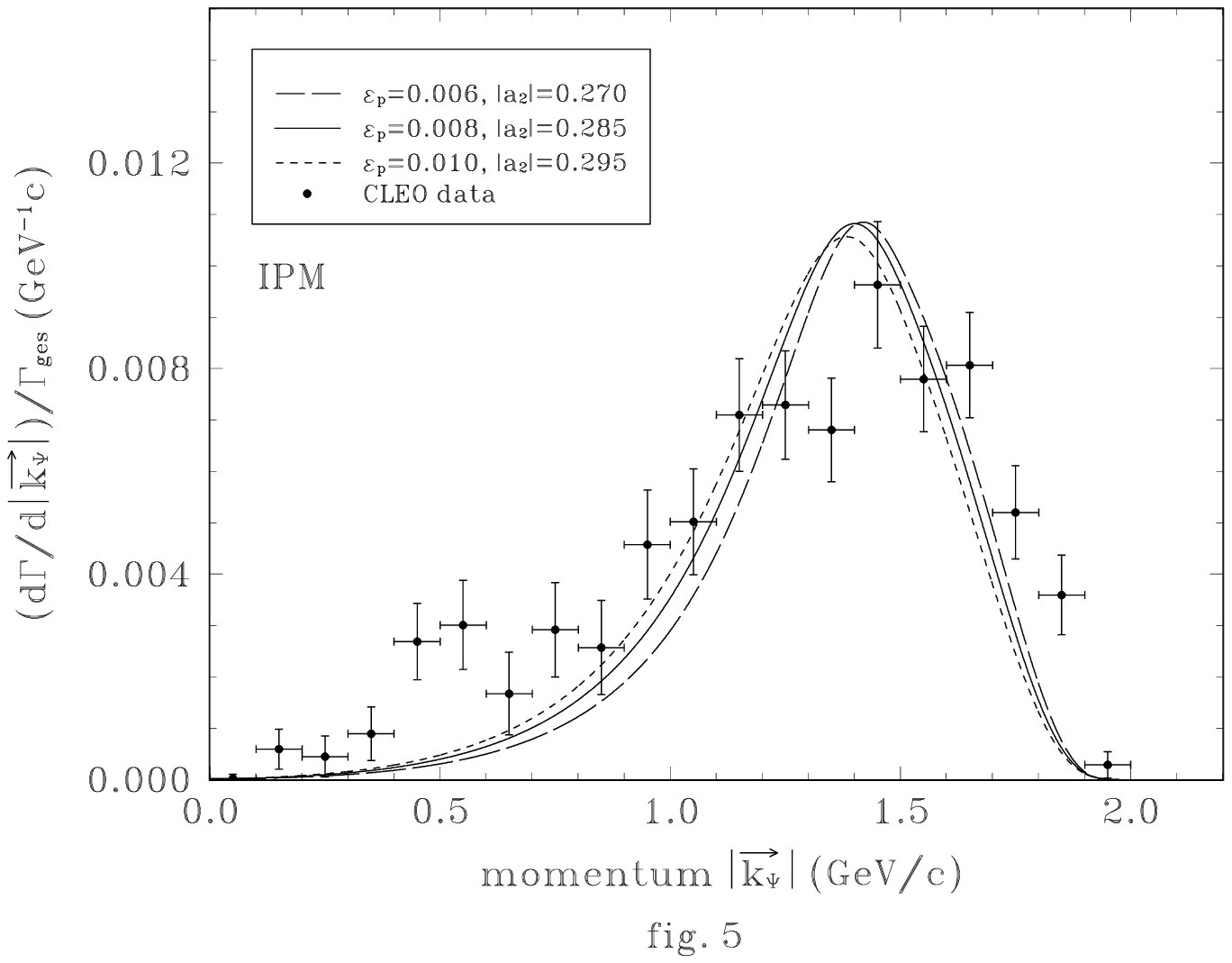,height=11cm,angle=360}\\
\vspace*{-1cm}
\noindent
\hspace*{1cm}\epsfig{file=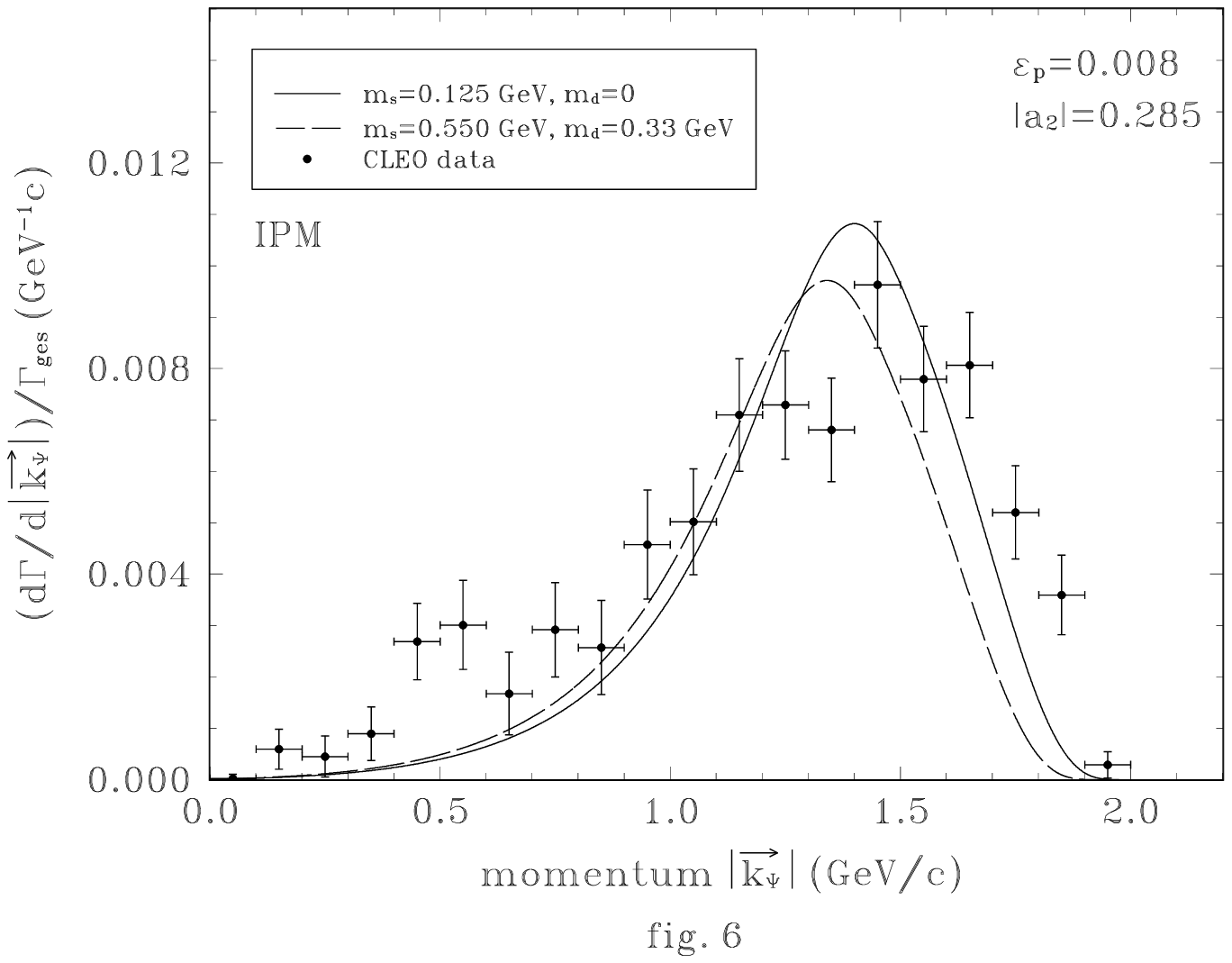,height=11cm,angle=360}
\newpage
\noindent
\hspace*{1.18cm}\epsfig{file=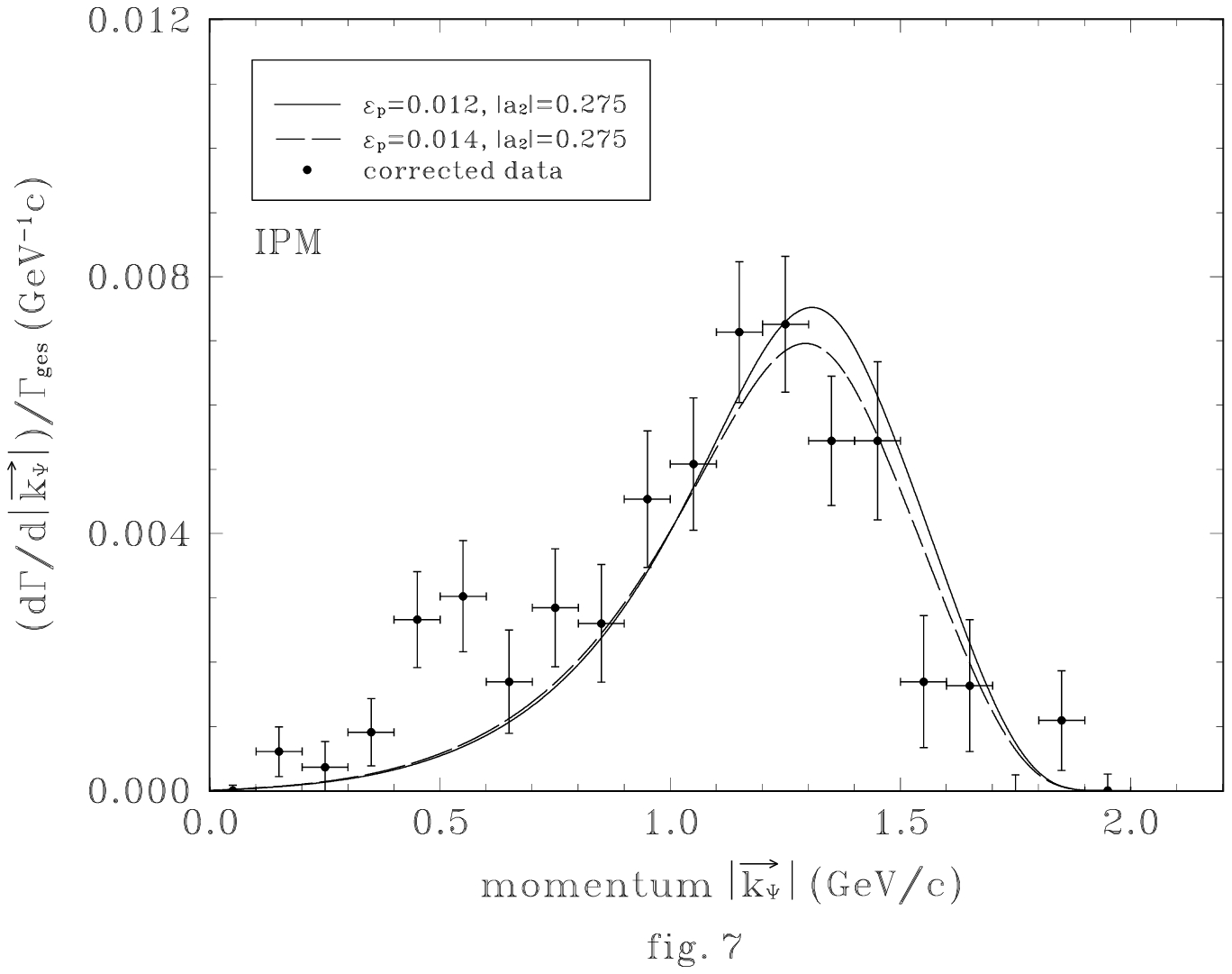,height=11cm,angle=360}\\
\vspace*{-1cm}
\noindent
\hspace*{1cm}\epsfig{file=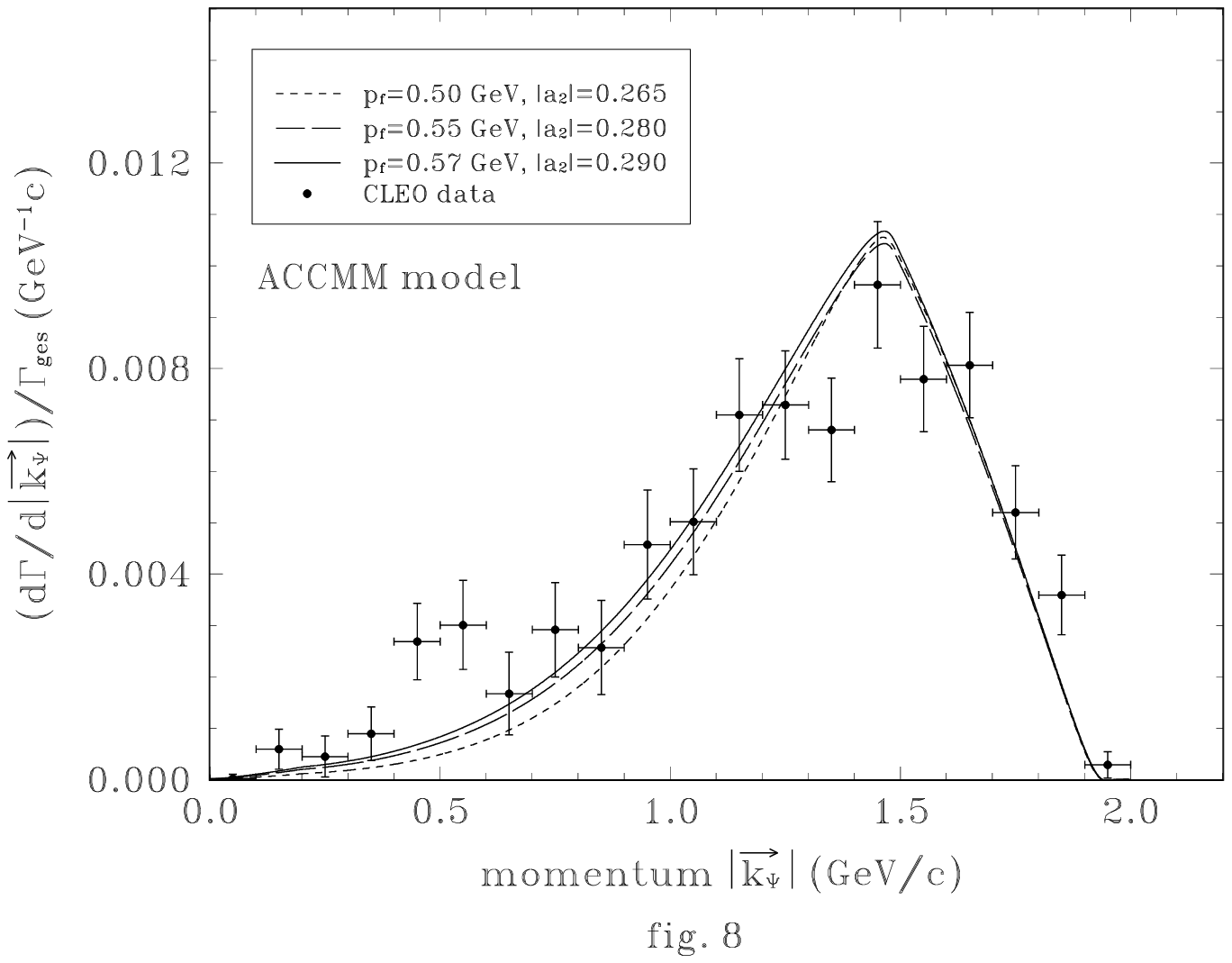,height=11cm,angle=360}
\newpage
\noindent
\hspace*{1.18cm}\epsfig{file=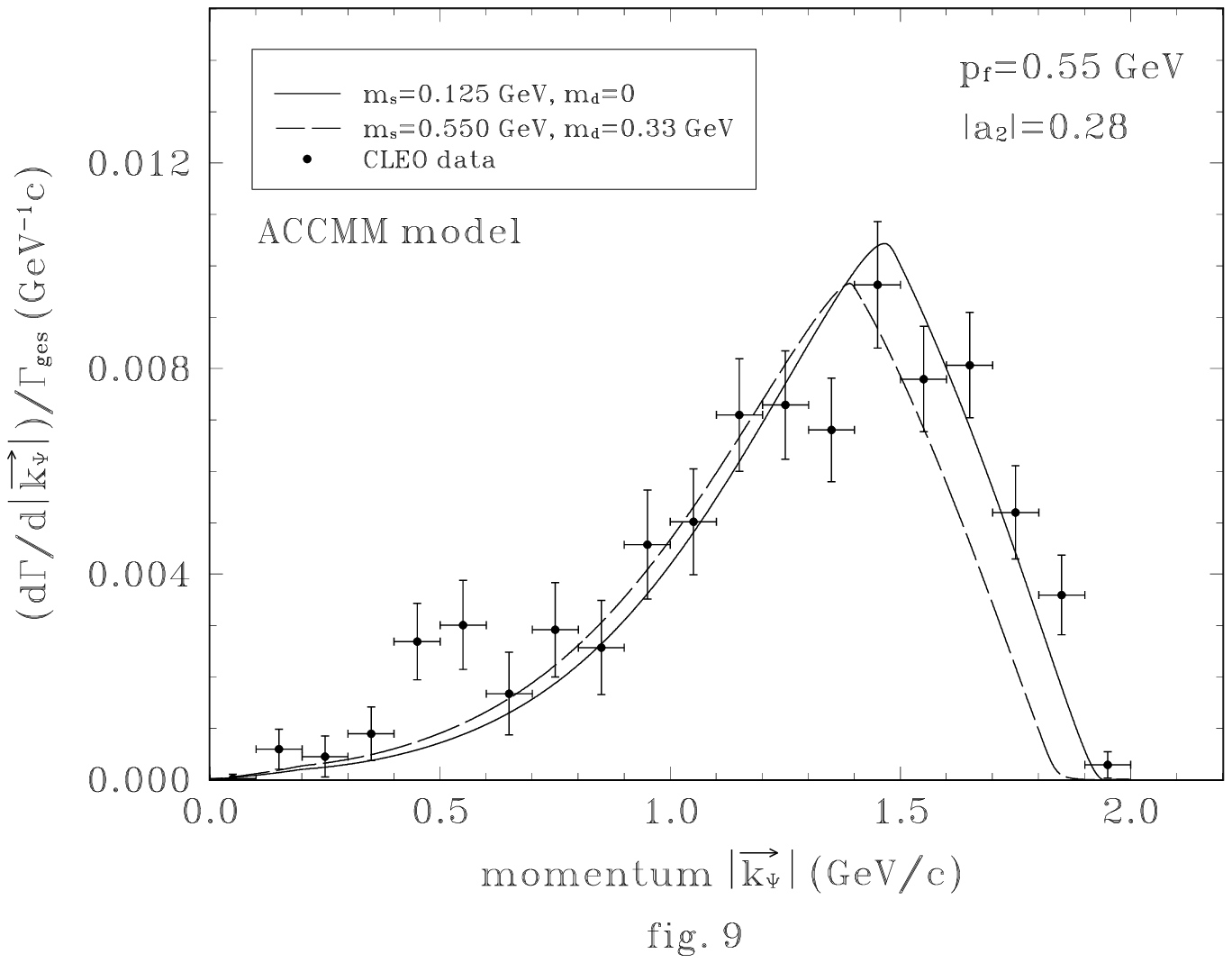,height=11cm,angle=360}\\
\vspace*{-1cm}
\noindent
\hspace*{1cm}\epsfig{file=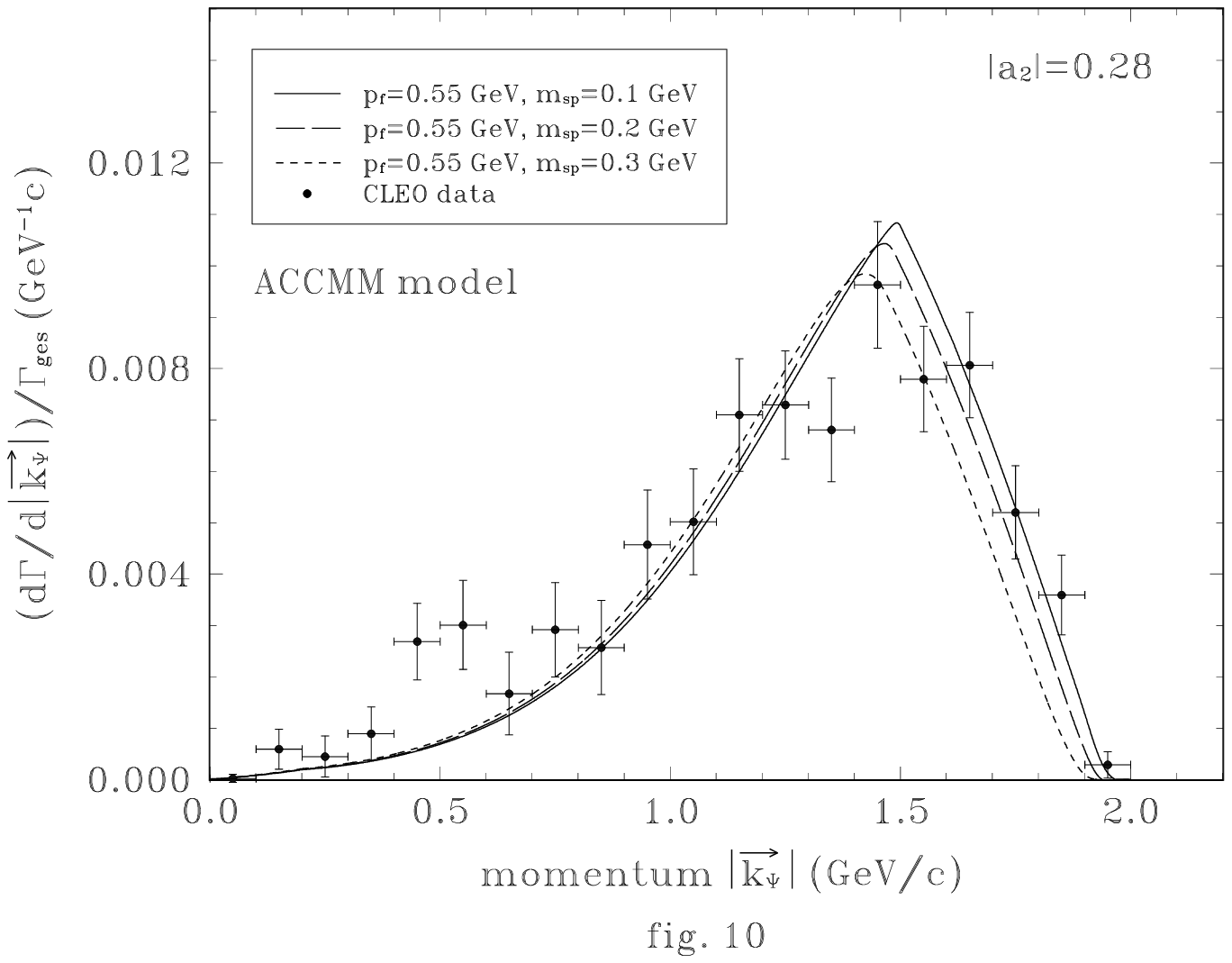,height=11cm,angle=360}
\newpage
\noindent
\hspace*{1.18cm}\epsfig{file=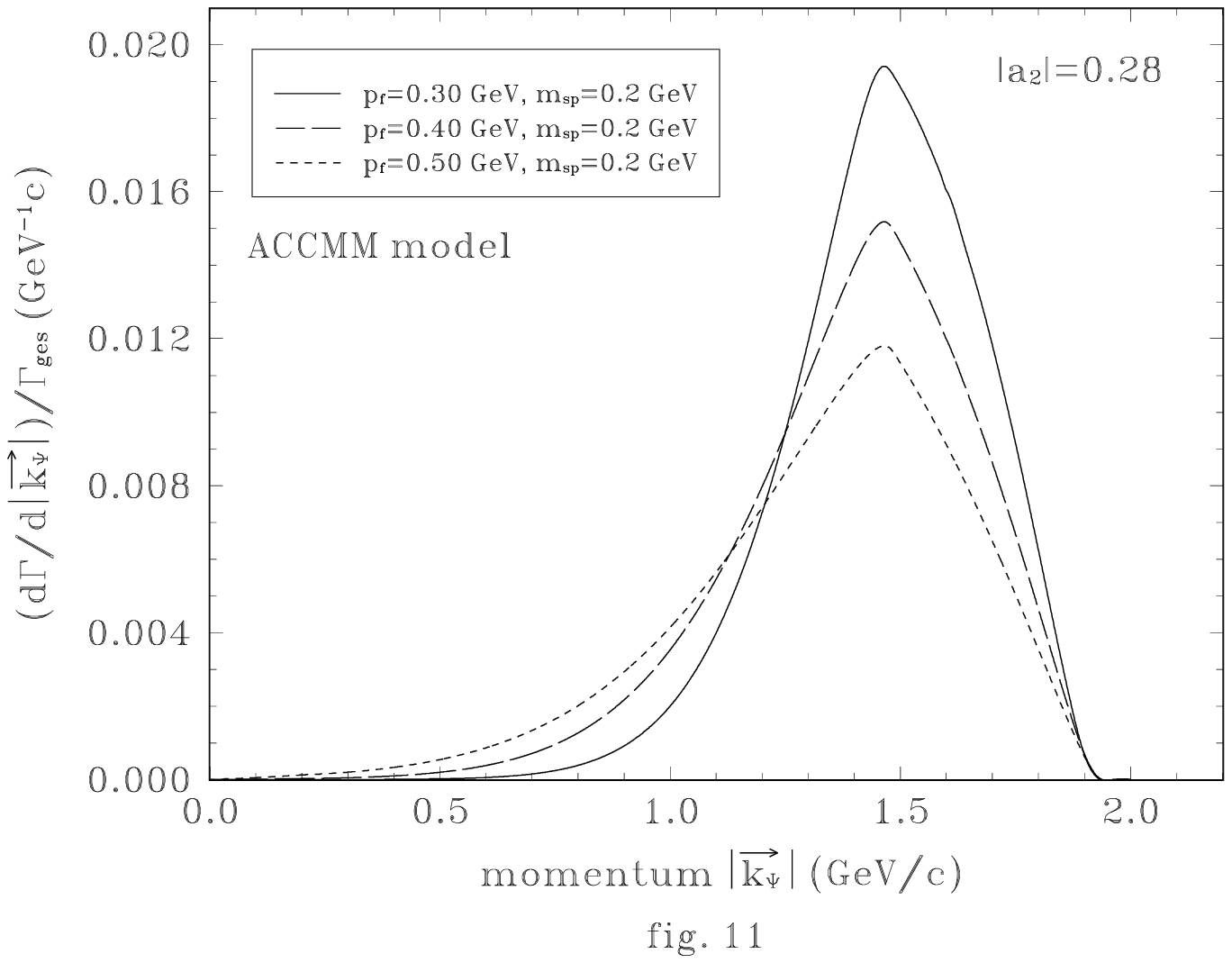,height=11cm,angle=360}\\
\end{document}